\documentclass[hidelinks,review]{elsarticle}

\usepackage{amsmath}
\usepackage{verbatim}
\usepackage{graphicx}
\usepackage{calc}
\usepackage{subfig}
\usepackage{textcomp}
\usepackage{verbatim}
\usepackage{longtable}
\usepackage{multirow}
\usepackage{lscape}
\usepackage{booktabs}

\usepackage[toc,page]{appendix}
\usepackage{chngcntr} 
\usepackage{dashbox} 

\usepackage{color}
\usepackage[usenames,dvipsnames,svgnames]{xcolor}

\usepackage[plainpages=false,pdfpagelabels]{hyperref}

\begin{document}

\begin{frontmatter}

\title{A Comparison and Catalog of Intrinsic Tumor Growth Models}

\author[hmc,focal]{E.A. ÷Sarapata\corref{cor2}}
\ead{esarapata@g.hmc.edu}

\author[hmc]{L.G. ÷de Pillis\corref{cor1}}
\ead{depillis@g.hmc.edu}

\cortext[cor1]{Corresponding author}
\cortext[cor2]{Principal corresponding author}
\fntext[fn1]{Telephone: 917-216-0573. This author has no affiliations or conflicts of interest.}

\address[hmc]{Department of Mathematics, Harvey Mudd College, Claremont, CA 91711}

\begin{abstract}

Determining the mathematical dynamics and associated parameter values that 
should be used to accurately 
reflect tumor growth 
continues to be of interest to mathematical modelers, experimentalists and practitioners. 
However, while there are several competing canonical tumor growth models that are 
often implemented, how to determine which of the models should be used for which tumor types
remains an open question. In this work, we 
determine the best fit growth dynamics and associated parameter ranges for
ten different tumor types 
by fitting growth functions to at least five sets of published 
experimental growth data per type of tumor. 
These time-series tumor growth data are used to determine 
which of the five most common tumor growth models 
(exponential, power law, logistic, Gompertz, or von Bertalanffy) 
provides the best fit for each type of tumor. 

\end{abstract}

\begin{keyword}
Population dynamics \sep Parameter fitting \sep Dynamical systems
\end{keyword}

\end{frontmatter}


\frontmatter

\section{Introduction}

Intrinsic tumor growth functions are a component of nearly all continuous, deterministic, cell-population based 
cancer models, yet there is no universal consensus as to which intrinsic growth function should be used
when a new mathematical model is being built. 
Of the published works which focus 
exclusively on intrinsic tumor growth, five models are widely used: exponential growth functions, power Law functions, 
logistic growth functions, von Bertalanffy growth functions, and Gompertz growth 
functions~\cite{depillis2006, agur1998}. In a 
study by Hart \emph{et al.}\cite{agur1998}, the authors compare 
Gompertz, logistic, exponential and power law growth against mammography data on human breast cancer.
The authors ultimately conclude that the power law should be used to represent breast cancer growth, 
though future 
investigations by the same authors found logistic growth to 
yield a better fit to the data of interest~\cite{agur1998, agur2011}. 
Another study by de Pillis and Radunskaya, in which intrinsic tumor growth functions for murine 
melanoma were compared, concluded that von Bertalanffy and logistic growth models provided the most accurate fit
to data~\cite{depillis2006}. 
A study by Zheng \emph{et al.}~\cite{zheng2013} 
compares exponential and biexponential models of lung cancer growth, 
in which the biexponential model is meant to approximate a tumor with two different 
speeds of growth. 
That study concludes that a biexponential model produces a better fit in all tested cases.

The choice of intrinsic growth function is strongly driven by the type of cancer being modeled, 
in addition to the environment in which it is growing ({\em e.g.}, in vitro, in mice, or in humans). In 
this work, we carry out a thorough exploration of a large collection of published cancer 
growth data in both mice and humans, and determine the best fit intrinsic growth 
functions along with the associated parameter ranges. For each of the ten tumor types we analyze, 
we have collected between five and ten separate sets of experimental data. After normalizing the data sets 
so they can be compared, we fit the data to exponential, power law, logistic, Gompertz and von Bertalanffy growth 
models. 

The process of comparing different growth functions to data 
naturally yields biologically relevant parameter values and ranges 
associated with each function.
We provide all those parameter ranges in this work.
Mathematical modelers must often be creative in choosing appropriate model parameters: they 
may borrow parameter values directly from published sources, or they may fit functions to data, 
if relevant data are available, or they may just have to use an {\em ad hoc} value, choosing a value 
that yields biologically reasonable dynamics. 
A large number of studies use experimental data from radically different experiments to estimate parameters; it is 
nearly unavoidable to combine data from murine and human sources, or obtained 
from \emph{in vitro} and \emph{in vivo} trials~\cite{jackson1999, depillis2013, tessi2012, sanga2006}.
The challenge of function choice and parameter determination will 
always be present for the modeler, and techniques for case-by-case parameter choice will still have to be 
pursued \cite{lillacci2010parameter}. However, the catalog of intrinsic growth laws and associated 
parameter ranges we provide for a variety of commonly modeled cancer types should provide a helpful starting 
point as researchers develop new models.

\section{Assumptions and Methods}

\subsection{Experimental data}

We curated 
time series tumor growth data sets for ten types of tumor: 
bladder cancer, breast cancer, colon cancer, head and neck squamous cell carcinoma, hepatocellular carcinoma, lung cancer, 
melanoma, ovarian cancer, pancreatic cancer, and renal cell carcinoma. 
Each group of data sets was collected from at least five peer-reviewed publications, with the smallest-sized group containing 
seven data sets and the largest containing seventeen data sets. 
In addition, at least one data set collected for each type of cancer was obtained 
from \emph{in vitro} trials and at least one data set was 
collected from \emph{in vivo} trials. 
Along with \emph{in vitro} trials, the range of target organisms included SCID mice, nude mice, normal mice, hamsters and humans. 
Table \ref{Tbl:CancersAndSources} 
shows all sources for each 
time series data set included in our study, as well as the cell lines for each trial.

\subsection{Unit normalization}

Among the publications that reported time series tumor growth data, 
the units and methods of tumor size measurement varied greatly. At least one paper per type of 
tumor was an \textit{in vitro} trial that reported tumor size as a cell number, the preferred unit for our purposes, 
but all data from \textit{in vivo} and \textit{in situ} trials were presented in units 
of mm$^3$, mm$^2$, mm, cm$^3$, or relative volume. A study by Dempsey \emph{et al.} demonstrated 
that unidimensional and bidimensional measures of tumor growth are less accurate as a predictor of survival than 
volumetric measures, lending to a possible source of pre-analytic error~\cite{dempsey2005measurement}.
In addition, instead of assuming a spherical tumor, volume was reported in a majority of papers as the product 
of the height, length and width of the tumor, overestimating the volume. However, we will also assume 
that no individual tumor cells are compressed, which will underestimate the number of tumor cells. 
The combination of these two assumptions is presumed to bring the estimated cell number within reasonable 
error of the real cell number.

In many cases, we were able to obtain an estimate of the number of tumor cells in a given volume from 
murine data sets that reported an initial cell count along with an initial volume measurement. We then divided the 
volume by the cell number, allowing for an estimate of the volume of a single tumor cell. We used this same estimate for 
data sets on tumor growth for tumors originating from the same organ. The most accurate conversion estimate, 
requiring the fewest conversions from the original data, was an estimate of $2.85\times 10^{3}$ cells$/\mu$m$^3$ for pancreatic 
cancer  \cite{kisfalvi2009}. For types of tumor that did not have a conversion data set available, we estimated the 
conversion ratio at approximately $1.82\times 10^{3}$ cells$/\mu$m$^3$~\cite{depillis2013}. Although these two 
estimates were obtained from different sources and for different cells, it should be noted that they are 
the same order of magnitude despite the high variability of cell size.

This volume estimate of a tumor cell provides a method with which to convert volume, area or length 
measurements to cell number. For those data sets which reported growth in volume, we normalized each 
datum by $\frac{1}{\mu_T}$ where $\mu_T$ is the tumor cell volume calculated as above. 
All of the publications that reported an 
area measurement obtained values by multiplying the minor axis 
of the tumor by the major axis  \cite{ricker2004, sunwoo2001, boukerche1989, juhl1997}. In this case, we assumed a cubic tumor with a volume of $a\sqrt{a}$ where $a$ is the reported area measurement. This allows us to calculate cell number from volume as before. Another set of papers reported only the major axis of the tumor  \cite{murgo1985, burke1997, fujimoto1995}. Here, we assumed a spherical tumor with the radius being one-half the major axis, using the volume of the sphere to estimate the cell number. For those papers that reported relative volume, we converted the data to cell number using the information in the supplemental material sections of each paper  \cite{okegawa2001, fujiwara1993, ahonen2000}.

\subsection{ODE Tumor Growth Models}

We compare fittings of tumor data for five different ODE growth models; exponential, power Law, logistic, Gompertz, and von Bertalanffy. Let $P$ represent an arbitrary population and let $t$ represent time. Exponential growth models are the simplest ODE growth model, described by
\begin{eqnarray}
\frac{dP}{dt} &=& rP
\end{eqnarray}
for some intrinsic growth rate constant $r$. Exponential growth is actually a special case of power law growth,
represented by
\begin{eqnarray}
\frac{dP}{dt} &=& rP^a,
\end{eqnarray}
where both $r$ and $a$ are parameters that must be fit to the data. Logistic growth, which incorporates a 
population carrying capacity, is given by
\begin{eqnarray}
\frac{dP}{dt} &=& rP\left(1-\frac{P}{K}\right)
\end{eqnarray}
where $r$ represents the intrinsic growth rate and $K$ represents the carrying capacity. 
Logistic growth looks very much like exponential exponential growth at low populations, 
but accounts for the resource-limited slowing of growth for larger populations.
Von Bertalanffy growth,  
also incorporating a carrying capacity, is given by
\begin{eqnarray}
\frac{dP}{dt} &=& r(K-P).
\end{eqnarray}
The final commonly used tumor growth model we will include is Gompertz growth, one form of which is given by
\begin{eqnarray}
\frac{dP}{dt} &=& r\log\left(\frac{K}{P}\right)P.
\end{eqnarray} 
Unfortunately, very few data sets track tumor growth long enough to 
sufficiently estimate carrying capacities. 
In order to get a good estimate, therefore,
we sought out data sets that recorded large 
tumor cell populations, and compared the former two models against the latter three  \cite{fujiwara1993, takahashi1992, murgo1985, richmond1983, kisfalvi2009, reinmuth2002, nakata1998, caltagirone2000, ricker2004}.


\subsection{Parameter fitting algorithms}

The parameters for each tumor growth model were estimated using at least two least-squares distance minimization algorithms. For each ODE model, the ODE with parameters was solved numerically using MATLAB's \verb=ode45= function, 
which adaptively implements a 4th or 5th order Runge-Kutta solver. 
We then minimized a least squares distance function between the numerical ODE solution and a target set of data 
using either MATLAB's built-in \verb=fminsearch= function or a Markov chain fitting with simulated annealing. 
MATLAB's \verb=fminsearch= is a Nelder-Mead simplex direct search function. Nelder-Mead is one of a class of 
local-search algorithms. Local algorithms require that the user provide an initial value sufficiently close to the 
sought after solution,
or the method may 
converge to nearby local minimum, but not to a global minimum. The local minimum found may not produce the 
best fit  \cite{lagarias1998}. This necessitates the use of an alternate global data fitting method.


The global data fitting method we implement is a Markov chain Monte Carlo (MCMC) fitting with simulated annealing, a non-deterministic search algortihm. MCMC evaluates a wider range of values in parameter space than does \verb=fminsearch=, and also includes a method for escaping from a local minimum to continue to search for a global minimum\citep{winkler2003, gilks1996, brooks2011}. The stochasticity in the algorithm yields different outcomes even among trials with the same initial conditions \citep{winkler2003, gilks1996, brooks2011}. The process is repeated $n$ times, where $n$ is an arbitrary number chosen by the user. The algorithm has no standard stopping condition. In our case, we chose $n = 200.$ This number of runs allowed us to achieve relatively good fits while keeping computational running times reasonable. Implementing the algorithm with a larger number of iterations $n$ may increase the chance that a global minimum is located.

Simulated annealing is the process of fitting, not to the distance function, but to the distance function raised to successive powers from 0 to 1, where the result of each fitting is used as the initial condition for the next fitting. The simulated annealing step reduces the chance that a minimization function will converge to a local minimum instead of a global minimum, since the act of raising the distance function to a power less than 1 reduces the prominence of local minima~\cite{winkler2003}. In our fittings, we ran 10 trials with simulated annealing, corresponding to ten iterations. Each iteration of the simulated annealing process involved $n$ repetitions of MCMC, where, as above, $n=200.$

While the local search algorithm \verb=fminsearch= will return
parameter values that produce the lowest least-squares fitting within a
bounded neighborhood of the initial parameters,  Markov chain methods such as MCMC
return the parameter values that produce the lowest least-squares fitting
over a finite number of arbitrary parameters from anywhere in the
parameter space\citep{ashyraliyev2009}. 
This difference in the domain
of each algorithm leads to defining behaviors that either help or hinder the
goodness of fit. The strength of \verb=fminsearch= 
is that will converge rapidly to a local minimum, as long as it is near one. However, 
it is is known to miss global minima that may produce a better fit. 
On the other hand, Markov chain methods like MCMC can locate minima that may be
far from the initial state, but they are less likely to hone in on the
exact minimum in a local sink.

In order to address the respective shortcomings of these global and local parameter
fitting algorithms, we used a hybrid approach that incorporates both Nelder-Mead simplex direct
search and Markov chain fitting with simulated annealing. We start
with one round of \verb=fminsearch= fitting. The resulting parameters are then 
passed as initial conditions to the MCMC algorithm.  MCMC is iterated a sufficient number of times
to yield parameters giving a good fit; in this case, 200 times.
Since MCMC is effective at breaking out of local minima and finding the neighborhood of a global minimum, but less effective
at actually converging to the minimum, a second round of \verb=fminsearch= is then performed,
using the results of the Markov chain fitting as initial conditions. This ensures
convergence to the deepest local minimum. All parameters reported in Section \ref{Sec:Results}
were determined using this sequence of fitting algorithms.


\subsection{Biologically Motivated Assumptions}

To determine the recommended parameter values for each growth function and each tumor type, 
we recorded the parameters of the function that best represented all trials with the same model organism at once. 
However, in order to determine appropriate parameter ranges, we performed fittings to each data set individually and recorded the extrema of each set of parameters. It is also assumed that \emph{in vitro} trials are better indicators of intrinsic tumor growth rates, due to the lack of an immune system in the growth environment; and that \emph{in vivo} trials are better indicators of animal carrying capacity, since the growth media are closer to conditions the tumor would encounter in a living organism. Thus, when relevant, intrinsic growth rates are determined from \emph{in vitro} trials only and carrying capacities are determined from \emph{in vivo} trials only. In cases where no carrying capacity is given, \emph{i.e.}, the exponential and power law growth models, only \textit{in vitro} trials are used to determine the growth rate, and the \textit{in vitro} trials are also used to determine the exponent for the power law model.

\subsection{Fitting evaluation metrics and Parameter sensitivity analysis}

We can compare the goodness-of-fit between the output of \verb=fminsearch= and MCMC by comparing 
the least-squares distances between each parameter-dependent function solution the data set to which it was fit. 
Using this metric, 
lower residuals (smaller least squares distances) suggest a better fit. We also use the least-squares residuals to 
calculate the Bayesian Information Criterion (BIC) for each fitting~\cite{o2004bayesian}.  The BIC
guards against over-fitting by accounting for goodness-of-fit while penalizing 
models that have larger numbers of parameters to be fit.

We carried out two types of parameter sensitivity analysis algorithms on the individual tumor growth models. 
A ``local'' or ``one-at-a-time'' parameter sensitivity analysis was performed to measure what the effect 
on the model outcome is when a single parameter is increased or decreased by some percentage of its value, while keeping other parameters constant.
We also carried out
a Partial Rank Correlation Coefficient test (PRCC), which 
is intended to measure the statistical influence on the model output of parameters that have monotonic but 
nonlinear behavior~\cite{boloye2012,iman1991repeatability,hamby1994review}. Since it is impossible to determine PRCC values from a model that has only one parameter, 
the exponential model is excluded from PRCC analysis.

A PRCC value close to zero implies that parameters are independent of one another.  
If the parameter space is large, Latin Hypercube Sampling can be used to provide input to the PRCC test by random sampling from 
an $n$-dimensional space for a model with $n$ parameter values~\cite{blower1994sensitivity, iman1988investigation, iman1980small}. These techniques are only applicable to models with more than one parameter; thus they are performed for the power law, logistic, Gompertz and von Bertalanffy models, but excluded for the exponential model.

\section{Results}\label{Sec:Results}

\subsection{Tumor Growth Parameter Values}

In order to determine a set of recommended parameters and appropriate ranges 
for each type of cancer and growth model, we fit the parameters of each growth equation to a minimum of five data sets per type of cancer. These parameters fall into three different classes: intrinsic growth rates (denoted $r$), exponents (denoted $a$) and carrying capacities (denoted $K$.) 
Two different types of fittings were performed on each set of related data sets. The \textit{in vitro} 
trials for each type of cancer were fitted separately for the best fit parameters to determine an 
acceptable parameter range, then together with different initial conditions to determine the 
recommended parameter values.

We provide a catalog of suggested parameter values and ranges for ten types of cancer 
and five models in Table \ref{t1}. 
The least squares residuals and BIC values for the combined fittings can be found in Table \ref{t2}. 
In order to highlight the best fits and the relationship between least squares residuals and BIC values, in each row,
the lowest least squares residuals values 
are outlined with solid borders, and BIC values are outlined with dashed borders. Graphs for each individual fitting and combined fittings, as well as the residuals, parameters and sources for all fittings, 
can also be found in \ref{App:1} through \ref{App:3}.
We were also able to determine a ranking of model fit for each cancer type from the evaluation metrics, shown in 
Table \ref{Tbl:ModelRanking}.  
This ranking was determined by comparing the sum of the least squares residuals for all individual and combined trials for each type of cancer.

\begin{table}
\begin{center}\hspace*{-2cm}
\begin{tabular}{lccccc}
\hline
\textbf{Cancer}& \multicolumn{5}{c}{\textbf{Model Ranking}}\\
& 1 & 2 & 3 & 4 & 5\\
\hline
Bladder & Power Law & Gompertz & Logistic & Exponential & Von Bertalanffy \\
Breast & Logistic & Gompertz & Power Law & Exponential & Von Bertalanffy\\
Colon & Power Law & Von Bertalanffy & Gompertz & Logistic & Exponential\\
HNSCC & Gompertz & Power Law & Exponential & Logistic & Von Bertalanffy\\
Liver & Logistic & Gompertz & Power Law & Von Bertalanffy & Exponential \\
Lung & Logistic & Power Law & Gompertz & Von Bertalanffy & Exponential\\
Melanoma & Power Law & Logistic & Exponential & Gompertz & Von Bertalanffy \\
Ovarian & Power Law & Exponential & Gompertz & Logistic & Von Bertalanffy \\
Pancreatic & Power Law & Gompertz & Logistic & Exponential & Von Bertalanffy\\
RCC & Power Law & Logistic & Exponential & Gompertz & Von Bertalanffy\\
\end{tabular}
\caption{Model Fit Ranking According to Least Squares Residuals}
\label{Tbl:ModelRanking}
\end{center}
\end{table}

\subsection{Parameter Sensitivity Analysis}

\begin{figure}
    \begin{center}
      \includegraphics[width=1.1\linewidth]{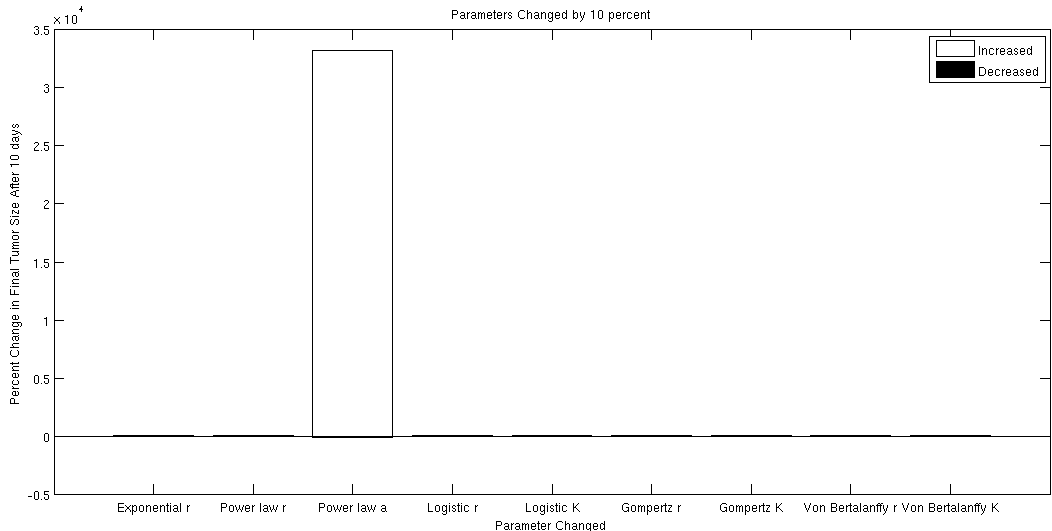}
    \end{center}
    \caption{Local Parameter Sensitivity Analysis for Five Models, Altering Parameters by 10\%}
    \label{Fig:SensWithPowerLaw}
  \end{figure}
\begin{figure}
    \begin{center}
      \includegraphics[width=1.1\linewidth]{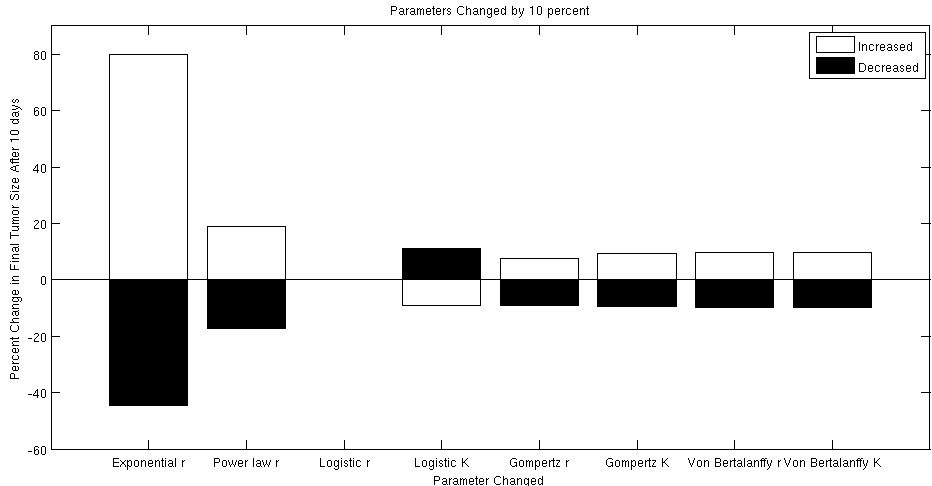}
    \end{center}
    \caption{Local Parameter Sensitivity Analysis for Five Models, Altering Parameters by 10\% (with power law $a$ results removed)}
     \label{Fig:SensWithoutPowerLaw}
  \end{figure}

The ``one-at-a-time'' parameter sensitivity analysis was carried out by 
altering each parameter by 10\%, with an initial condition of $1\times 10^4$ tumor cells, 
running the model for 10 days, and starting with the parameters from the individually 
determined \textit{in vitro} colon trials. The results are presented in Figure \ref{Fig:SensWithPowerLaw} and 
Figure \ref{Fig:SensWithoutPowerLaw} (where Figure \ref{Fig:SensWithoutPowerLaw} has the power law exponent removed to increase readability of the percent changes associated with the other parameters.) We also provide a PRCC analysis over 1000 randomized parameter values using Latin hypercube sampling, which is presented in Table \ref{Tbl:PRCC}.

\begin{table}
\begin{center}
\begin{tabular}{cc}
\hline
Parameter & PRCC\\
\hline
Power law $r$ & 0.0412\\
Power law $a$ & 0 \\
Logistic $r$ & 0\\
Logistic $K$ & 0\\
Gompertz $K$ & 0.0292\\
Gompertz $K$ & 0\\
Von Bertalanffy $K$ & - 0.0104\\
Von Bertalanffy $K$ & 0\\
\end{tabular}
\caption{Results of Partial Rank Correlation Coefficient Test for Two-Parameter Growth Models}
\label{Tbl:PRCC}
\end{center}
\end{table}

\section{Discussion}

\subsection{Model Comparison}

Table \ref{Tbl:ModelRanking} provides a summary of 
the best fit models for each type of tumor. 
These results suggest that of the models tested, there is no one model that best approximates all forms of tumor growth. 
The power law provides especially close fits to data that do not appear to approach a carrying capacity, most likely because
it is more flexible in approximating exponential growth dynamics. 
The Gompertz and logistic models outperform either the von Bertalanffy or exponential models in each case.

In some cases, the results of the fitting algorithm may be misleading. Logistic growth fittings sometimes resulted in
a carrying capacity with an order of magnitude much higher than comparable trials but with the same intrinsic growth rate as the 
exponential fit to the same data. 
This occurs with \textit{in vivo} trial 3 for breast cancer;  \textit{in vivo} trial 4 for head and neck squamous cell carcinoma; 
\textit{in vitro} trials 1, 2, and 10 and \textit{in vivo} trial 1 and the combined \textit{in vivo} fit for lung cancer; 
and \textit{in vitro} trials 1 and 3 for ovarian cancer. 
When this happens, it may be that the exponential fit is a better match to the data than the logistic fit. 
In such a case, the logistic growth function may be approximating exponential growth by raising the carrying capacity to a 
number high enough so that it does not affect the fitting. This theory is supported by the least squares residuals; 
the least squares residuals from the exponential fit and the residuals from the logistic fit are the same when this situation occurs.

One concern that must be addressed is whether the best-fit parameters are biologically accurate \cite{slezak2010}. We note that the best-fit von Bertalanffy parameters, which are expected to have intrinsic growth rates similar to all other models, consistently have intrinsic growth rates that are two or three orders of magnitude smaller. This is enough of an indication to doubt the biological accuracy 
of the von Bertalanffy parameters obtained by least-squares fitting. In addition, we have reason to question the biological relevance of the power law fittings for similar reasons. 

\subsection{Concerns about Power Law}

We conclude from the parameter fitting process that it may not be justifiable to alter power law growth parameters, even within the range given by repeated fits. This is because the best fit power law parameters occasionally have uncharacteristically high intrinsic growth rates (e.g. \textit{in vivo} breast cancer trials 1 and 2, the combined head and neck squamous cell carcinoma \textit{in vivo} trial, \textit{in vitro} lung trial 5) and exponents that are lower than the exponents in trials in the same cancer. These results suggest that power law fitting is highly sensitive, where the intrinsic growth rates rise unpredictably to accommodate lower exponents and vice versa. Therefore, although power law fits occasionally have lower residuals than the other growth laws, their unstable nature would
prevent modelers from changing parameters even slightly within a specified range. 

In addition, the sensitivity analyses can be used to provide a basis for our claim that the power law is not a viable model. 
Figure \ref{Fig:SensWithPowerLaw} suggests that $a$, the exponential component of the power law model, affects the model output at a much higher percentage than any other parameter in any other model. In fact, increasing $a$ by only 10\% caused the tumor to grow almost 35000\% larger in only 10 days. This suggests that altering $a$ individually would change the tumor growth behavior at a massive rate that has no biological justification. An alternative would be to alter $a$ and $r$ in conjunction, such that the relatively low least squares residuals for the fitting are preserved. However, as the PRCC results suggest, the relationship between $a$ and $r$ is highly nonlinear. This is not suggestive in and of itself---none of the other parameters had significant PRCC results---rather, we draw the conclusion in light of the results of the ``one-at-a-time'' 
parameter sensitivity analysis. 
In practice, a researcher seeking to lower the growth rate or raise the exponent of some of the less biologically sound power law fittings would have difficulty determining a relationship between $a$ and $r$ that allows the parameters to be altered while preserving the behavior of the original curve. This rigidity and extreme sensitivity is what makes the power law a less than ideal choice for a tumor growth model.

For these reasons, despite the low residual fits we found, we discourage 
the use of the power law model.

\subsection{Parameter Fitting Algorithms}

To see that the hybrid fitting algorithm is more effective than either the Nelder-Mead simplex direct search or Markov Chain method with simulated annealing, we note that a set of parameters is only accepted if the least squares residuals are lower than they were in the previous fitting. Since \verb=fminsearch= is used to provide initial values for the Markov Chain method, the residuals of a Nelder-Mead simplex direct search on a given data set bound the residuals of the hybrid search from above. Due to the nondeterministic nature of the Markov Chain method, the residuals are not necessarily always greater than those of the hybrid method, but it is true that the residuals returned by a specific iteration of the Markov Chain method will always be greater than the results of the hybrid algorithm using that specific iteration of the Markov Chain method. We have noted the inability of the Markov chain method to converge on global minima.


One may wonder whether
using a hybrid fitting algorithm than is necessary when \verb=fminsearch= may have been sufficient. 
One issue with \verb=fminsearch= is the inability to converge to a better minimum once a local minimum is 
detected by the algorithm, and to improve the fitting would necessitate changing parameters by hand. Since 
this project required 20 separate parameter fitting trials each to 70 data sets, 
not including the 90 combined fittings, manually altering parameters was not a viable option. Thus, 
even one instance of \verb=fminsearch= converging to a non-global minimum would necessitate the use of a 
stronger parameter fitting algorithm. This hybrid method was adopted after 
repeated difficulties with \verb=fminsearch= which would have remained unfixable otherwise.

As a side note, it is possible to fit the equations with carrying capacities in two different ways: the first, defining the parameters to be estimated as $r$ and $K$, and the second, defining the parameters to be estimated as $r$ and $1/K$. Although theoretically equivalent, these two approaches can produce different outcomes depending on which fitting metric is used. For the logistic equation defined as 
\begin{eqnarray}
\label{eq:StandardLog}
\frac{dP}{dt} &=& rP\left(1-\frac{P}{K}\right),
\end{eqnarray}
it is possible that the fitting algorithm may be slower in converging to the best-fit $K$, because it is possible for the best-fit $K$ to be several orders of magnitude higher than the initial condition. However, for the logistic equation defined as 
\begin{eqnarray}
\frac{dP}{dt} &=& rP\left(1-bP\right)
\end{eqnarray}
where $b=\frac{1}{K}$, both the Nelder-Mead simplex direct search and the MCMC 
method occasionally produced results where the carrying capacity was negative. This is a result of the relative distance in parameter space from the negative real axis; $1\times 10^6$ and $-1\times 10^6$ are much further apart than are $1\times 10^{-6}$ and $-1\times 10^{-6}$, for example. 
Therefore,  fitting to equation \eqref{eq:StandardLog} makes it more difficult for either algorithm to reach negative values. 
We therefore recommend fitting logistic growth using the parameter forms of equation \eqref{eq:StandardLog} 
in order to avoid the fittings from producing a biologically inaccurate carrying capacity.

\subsection{Parameter Sensitivity Analysis}

While it may seem odd to perform a sensitivity analysis on a series of models that each have only one or two parameters, these techniques can be interpreted to compare the justifiability of modifying parameters in each growth model. The PRCC values provide a measure of the strength of the relationship between two parameters, while the ``one-at-a-time'' parameter sensitivity analysis measures the effect of individual 
parameters on the model output. Therefore, while the ``one-at-a-time'' parameter sensitivity analysis can be used to 
estimate the effects of changing the value of a single parameter, the PRCC measure can tell us whether altering a single parameter while leaving the other constant is justifiable. If the ``one-at-a-time'' parameter sensitivity analysis reveals that altering a parameter by 
a small amount changes the output of the model by a significant amount, 
then researchers should be careful when modifying these parameters. 

\textbf{Acknowledgements}

We would like to thank Harvey Mudd College for providing us with the resources to complete this publication, and Ami Radunskaya, the second reader.\\

\textbf{References}

\bibliographystyle{elsarticle/model1-num-names}
\bibliography{bib2}

\appendix
\counterwithin{figure}{section}
\counterwithin{table}{section}


\section{Supplemental Materials: Tumor Growth Parameters} \label{App:1}


We present a catalog of suggested parameter values and ranges for ten types of cancer and the five canonical growth functions that were considered. The parameters found using a hybrid fitting algorithm to a minimum of five data sets per type of tumor are given in Table \ref{t1}. For comparison purposes, the least squares residuals and BIC values are presented in Table \ref{t2}. We highlight the lowest least squares residuals in each row with a solid border and the lowest BIC values in each row with a dashed border.

\begin{landscape}
\begin{table}
  \fontsize{8.3pt}{12pt}\selectfont
    \begin{center}
    \hspace{-2.5cm}
      \begin{tabular}{llllllll}
        \toprule
        Cancer & & Exponential & Power Law & Logistic & Gompertz & Von Bertalanffy \\
        \midrule
        \multirow{2}{*}{Bladder} & r & 0.0165:(0.0942):0.1919 & 0.0033:(5.5976):51.2297 & 0.1378:(1.3454):9.0473 & 0.0075:(0.2893):0.2893 & 3.6E$-4$:(0.0392):0.0037         \\
        & a, K& & 0.6839:(0.8582):1.1552 & 5.36E4:(1.24E9):1.91E10 & 3.07E5:(8.10E6):9.01E11 & 1.5E4:(1.5E4):1.11E11 \\       
        \multirow{2}{*}{Breast} & r & 0.183:(0.4593):1.3311 & 0.0456:(5.3077):3988.3 & 0.183:(1.4808):1.5488 & 0.0095:(0.4834):0.4834 & 6.17E$-5$:(2.95E$-4$):0.0062     \\
        & a, K& & 0.3151:(0.8299):1.122 & 9.45E5:(2.46E9):2.73E9 & 1.92E6:(1.53E10):4.24E13 & 3.95E8:(1.06E10):5.2E11 \\    
        \multirow{2}{*}{Colon} & r & 0.4555:(0.4816):0.5533 & 1.606:(1.606):35.339 & 0.5521:(0.5775):0.8401& 0.0608:(0.0608):0.2405 & 1.8E$-4$:(1.8E$-4$):2.6E4        \\
        & a, K& & 0.5521:(0.8819):0.8819 & 2.32E8:(8.02E8):3.65E51.47E9 & 3.78E8:(9.90E8):1.44E12 & 1.65E9:(4.49E10):2.04E11 \\    
        \multirow{2}{*}{HNSCC} & r & 0.0277:(0.0286):0.5014 & 0.0728:(0.6353):42.0098 & 0.0309:(0.0328):0.576 & 0.0017:(0.0044):0.1218 & 1.19E$-5$:(1.19E$-5$):1.51E$-4$  \\
        & a, K& & 0.7384:(0.8376):0.9597 & 5.17E6:(1.37E7):1.37E7 & 1.11E7:(1.17E12):7.01E14 & 7.66E8:(2.94E10):2.94E10 \\   
        \multirow{2}{*}{Liver} & r & 0.1749:(0.1754):0.5307 & :2.8332(278.3617):738.1756 & 0.2402:(0.2421):0.5982 & 0.0637:(0.0670):0.0730 & 6.99E$-5$:(6.99E$-5$):2.53E$-4$  \\
        & a, K& &0.5685:(0.619):0.8589 & 1.34E6:(4.85E8):2.44E9 & 3.04E8:(4.96R8):1.60E13 & 2.51E7:(5.01E8):3.55E10 \\     
        \multirow{2}{*}{Lung} & r & 0.0804:(0.3358):0.6501 & 0.0026:(0.2814):2.15E6 & 0.35:(0.381):1.3577 & 0.004:(0.0049):0.658 & 2.93E$-5$:(2.93E$-5$):0.1143       \\
        & a, K& & -0.0806:(0.995):1.3872 & 3.77E7:(1.36E12):1.36E12 & 3.85E7:(2.84E21):2.84E21 & 4.02E7:(9.70E11):2.07E12 \\   
        \multirow{2}{*}{Melanoma} & r & 0.0908:(0.1502):0.2414 & 0.0081:(0.0081):299.271 & 0.1061:(0.1502):0.4766  & 0.0043:(0.0043):0.2872 & 0.0015:(0.5303):0.5303           \\
        & a, K& & 0.5442:(1.1596):1.1596 & 4.87E7:(7.36E8):2.36E9 & 7.93E7:(5.02E8):3.48E9 & 9.48E7:(4.06E8):3.49E11\\       
        \multirow{2}{*}{Ovarian} & r & 0.5087:(0.6765):0.7634 & 1.42E$-4$:(2.3356):3.58E7 & 0.5087:(0.823):2.4545 & 0.0193:(0.0815):1.4489 & 3.35E$-4$:(4.96E$-4$):0.7245     \\
        & a, K& & -0.4252:(0.8963):1.7012 & 8.08E8:(3.07E12):3.07E12 & 2.06E9:(9.41E19):9.41E19 & 9.32E10:(1.20E13):1.20E13 \\  
        \multirow{2}{*}{Pancreatic} & r & 0.0541:(0.3093):0.4122 & 2.84E$-5$:(0.668):13.216 & 0.0541:(0.348):0.4645 & 0.0023:(0.0524):0.1566 & 1.92E$-5$:(0.0055):0.0055        \\
        & a, K& & 0.7126:(0.9102):1.493 & 5.69E7:(5.69E7):7.62E8 & 7.09E7:(7.09E7):5.81E15 & 1.79E9:(6.28E9):2.81E11 \\     
        \multirow{2}{*}{RCC} & r & 0.3674:(0.4504):0.5568 & 0.6554:(44.7253):172.3193 & 0.3884:(0.5719):1.365 & 0.0342:(0.0773):0.8647 & 3.5E$-6$:(3.5E$-6$):0.4437       \\
        & a, K& & 0.2805:(0.6182):0.9473 & 2.49E8:(8.91E8):1.41E9 & 3.02E8:(3.02E8):4.19E11 & 2.95E8:(2.95E8):1.46E11 \\    
        \bottomrule
      \end{tabular}
      \caption{Recommended Parameter Values and Ranges for Ten
        Different Types of Cancer and Five ODE Growth Laws}
       \label{t1}
    \end{center}
\end{table}
  \end{landscape}

\begin{landscape}
\begin{table}
\fontsize{10pt}{12pt}\selectfont
\begin{center}
 \hspace{-3cm}
\begin{tabular}{lcccccccccc}
\toprule
Trial & \multicolumn{2}{c}{Exponential} & \multicolumn{2}{c}{Power Law} & \multicolumn{2}{c}{Logistic} & \multicolumn{2}{c}{Gompertz} & \multicolumn{2}{c}{Von Bertalanffy}\\
\cmidrule(r){2-3} \cmidrule(rl){4-5} \cmidrule(l){6-7} \cmidrule(l){8-9} \cmidrule(l){10-11}
& Residuals & BIC& Residuals & BIC& Residuals & BIC & Residuals & BIC & Residuals & BIC\\
\midrule
Bladder \textit{in vitro} & 7.43E10 & 97.117 & 4.22E10 & 96.241 & 3.00E10 & 94.876 & \fcolorbox{black}{white}{3.30E8} & \dbox{76.837} & 2.40E11 &103.194\\
Bladder \textit{in vivo} & 7.43E10 & 2363.3 & 4.22E10 & 2339.9 & \fcolorbox{black}{white}{3.00E10} & 2353.2 & 2.21E17 & \dbox{2329.9} & 1.09E18 & 2337.0\\
Breast \textit{in vitro} & 5.70E12 & 506.07 & \fcolorbox{black}{white}{2.04E12} & \dbox{489.52} & 2.28E12 & 491.59 & 2.08E12 & 489.87 & 2.39E12 & 492.55\\
Breast \textit{in vivo} & 2.10E18 & 749.61 & 1.40E18 & 744.80 & \fcolorbox{black}{white}{6.02E17} & \dbox{728.80} & 1.01E18 & 738.68 & 8.92E18 & 780.03\\
Colon \textit{in vitro} & 1.72E9 & 228.87 & 1.18E9 & 226.87 & 1.18E9 & 226.88 & \fcolorbox{black}{white}{1.18E9} & \dbox{226.83} & 1.03E10 & 252.80\\
Colon \textit{in vivo} & 1.79E18 & 1799.3 & 8.71E17 & 1769.3 & 1.33E18 & 1789.1 & 1.12E18 & 1781.1 & \fcolorbox{black}{white}{8.60E17} & \dbox{1768.6}\\
HNSCC \textit{in vitro} & 6.02E17 & 1531.0 & \fcolorbox{black}{white}{5.00E17} & \dbox{1527.0} & 5.57E17 & 1532.4 & 5.09E17 & 1527.8 & 1.17E18 & 1562.1\\
HNSCC \textit{in vivo} & 5.03E13 & 519.78 & \fcolorbox{black}{white}{3.07E13} & \dbox{513.78} & 3.77E13 & 517.48 & 5.13E13 & 523.02 & 2.30E14 & 550.03\\
Liver \textit{in vitro} & 1.51E17 & 770.77 & 2.15E16 & 732.92 & \fcolorbox{black}{white}{9.64E15} & \dbox{716.08} & 1.59E16 & 726.60 & 3.60E17 & 792.09\\
Liver \textit{in vivo} & 1.17E19 & 1809.5 & 8.60E18 & 1799.3 & \fcolorbox{black}{white}{7.51E18} & \dbox{1793.2} & 7.56E18 & 1793.4 & 7.83E18 & 1795.0\\
Lung \textit{in vitro} & 1.75E22 & 4201.0 & 1.74E22 & 4214.2 & \fcolorbox{black}{white}{1.29E22} & \dbox{4187.3} & 2.14E22 & 4232.6 & 1.75E22 & 4214.6\\
Lung \textit{in vivo} & 1.86E18 & \dbox{861.64} & \fcolorbox{black}{white}{1.64E18} & 861.85 & 1.86E18 & 864.73 & 1.90E18 & 865.11 & 2.86E18 & 874.11\\
Melanoma \textit{in vitro} & 5.25E18 & 688.49 & \fcolorbox{black}{white}{1.26E18} & \dbox{667.04} & 5.25E18 & 691.32 & 5.67E18 & 692.62 & 7.99E18 & 698.45\\
Melanoma \textit{in vivo} & 7.20E18 & 1903.2 & \fcolorbox{black}{white}{2.42E18} & \dbox{1854.8} & 6.48E18 & 1902.0 & 7.67E18 & 1910.1 & 8.34E18 & 1914.2\\
Ovarian \textit{in vitro} & 1.60E11 & \dbox{372.25} & 1.58E11 & 374.82 & \fcolorbox{black}{white}{1.56E11} & 374.55 & 1.58E11 & 374.80 & 2.20E11 & 380.09\\
Ovarian \textit{in vivo} & 3.27E18 & \dbox{758.01} & \fcolorbox{black}{white}{2.88E18} & 758.54 & 3.27E18 & 760.95 & 7.45E18 & 788.85 & 3.44E20 & 849.44\\
Pancreatic \textit{in vitro} & 5.87E5 & \dbox{61.088} & 5.86E5 & 62.534 & \fcolorbox{black}{white}{5.48E5} & 62.359 & 5.79E5 & 62.630 & 2.09E6 & 69.045\\
Pancreatic \textit{in vivo} & 5.79E16 & 1401.1 & \fcolorbox{black}{white}{3.83E16} & \dbox{1388.2} & 4.50E16 & 1394.7 & 4.15E16 & 1391.4 & 3.83E16 & \dbox{1388.2}\\
RCC \textit{in vitro} & 6.90E11 & 1140.5 & 6.40E11 & 1138.9 & \fcolorbox{black}{white}{4.71E11} & 1131.3 & 4.92E11 & \dbox{668.17} & 1.12E12 & 691.14\\
RCC \textit{in vivo} & 1.59E18 & \dbox{2052.6}& \fcolorbox{black}{white}{1.51E18} & 2054.0 & 1.59E18 & 2056.6 & 1.80E18 & 2063.5 & 1.83E18 & 2064.4\\
\bottomrule
\end{tabular}
\caption{Model Evaluation Metrics for Combined Experimental Data Fittings}
\label{t2}
\end{center}
\end{table}
\end{landscape}

\section{Supplemental Materials: Sources of Data for Parameter Values}\label{App:2}


A large number of individual studies were gathered in determining appropriate timescale tumor growth data sets to be used in the fitting process. Not only are the sources for each type of cancer listed, the individual cell lines used in each paper are included for posterity. Some papers, which used tissue samples from human subjects as the source of cancerous cells, did not specify a cell line.

\begin{longtable}{ll}
\toprule
Cancer and Cell Line & Sources\\
\midrule
\endhead
\bottomrule
\caption{Sources of Timescale Data by Type of Cancer and Cell Line}
\endfoot
\label{Tbl:CancersAndSources}
\textbf{Bladder Cancer} & \\
HT1376 & \cite{golshani2008}\\
UMUC-3 &  \cite{kamada2007}\\
KoTCC-1 &  \cite{miyake2001}\\
EJ-1 & \cite{ohnishi2003, du2003}\\
\textbf{Breast Cancer} & \\
MDA-MB-435BAG & \cite{coopman2000}\\
MCF-7&  \cite{Lu1999} \\
KPL-1 &  \cite{Nakagawa2001}\\
4T1-GFP-FL &  \cite{smith2004}\\
\textbf{Colon Cancer} & \\
KM12L4 & \cite{reinmuth2002}\\
Moser &  \cite{sarraf1998}\\
HCT116 &  \cite{sarraf1998,sheng1997}\\
CX-1 &  \cite{sarraf1998}\\
HCA7 &  \cite{sheng1997}\\
LS LiM6&  \cite{ warren1995}\\
Unspecified &  \cite{todaro2007}\\
\textbf{Head and Neck Squamous Cell Carcinoma} & \\
UM-SCC-9 &  \cite{duffey1999}\\
Tu-138 &  \cite{liu1999}\\
Tu-167 &  \cite{liu1999}\\
686LN &  \cite{liu1999}\\
CAL27&  \cite{lotempio2005}\\
UM-SCC-X &  \cite{ricker2004}\\
PAM-LY2 &  \cite{sunwoo2001}\\
\textbf{Hepatocellular Carcinoma} & \\
HCC-26-1004 &  \cite{huynh2008}\\
HCC-2-1318 &  \cite{huynh2008}\\
SH-J1 &  \cite{jung2006}\\
PLC &  \cite{liu2005}\\
Hep3B &  \cite{liu2005}\\
SMMC-7721 &  \cite{wong2005}\\
Unspecified &  \cite{zender2008}\\
\textbf{Lung Cancer} & \\
SW-900 &  \cite{esquelakerscher2008}\\
H226 &  \cite{esquelakerscher2008}\\
A549 &  \cite{esquelakerscher2008}\\
& \cite{fabbri2005}\\
&  \cite{ tsubouchi2000}\\
H460 &  \cite{fabbri2005}\\
H1299 &  \cite{fabbri2005}\\
U2020 &  \cite{fabbri2005}\\
H322a &  \cite{fujiwara1993}\\
WT226b &  \cite{fujiwara1993}\\
NCI-H727 &  \cite{moody1993}\\
3LL &  \cite{ sharma1999}\\
NCI-H358 &  \cite{takahashi1992}\\
H841 &  \cite{ tsubouchi2000}\\
pc14 &  \cite{ tsubouchi2000}\\
\textbf{Melanoma} & \\
M3Dau &  \cite{boukerche1989}\\
MIRW5 &  \cite{bregman1986}\\
B16-BL6 &  \cite{caltagirone2000, murgo1985}\\
A-375 &  \cite{kuntsfeld2003}\\
M21 &  \cite{petitclerc1999}\\
Hs0294 &  \cite{richmond1983}\\
Unspecified &  \cite{abe2004}\\
\textbf{Ovarian Cancer} & \\
SKOV-3 &  \cite{juhl1997, polato2005}\\
HRA &  \cite{nakata1998}\\
A2780 &  \cite{polato2005}\\
IGROV-1 &  \cite{polato2005}\\
HCT-116 &  \cite{polato2005}\\
MA148 &  \cite{yokoyama2000}\\
\textbf{Pancreatic Cancer} & \\
PC-1 &  \cite{burke1997}\\
MIAPaCa-2 &  \cite{ito1996, kisfalvi2009}\\
PANC-1 &  \cite{kisfalvi2009}\\
PancTu1 &  \cite{volger2008}\\
HPAC &  \cite{zervos1997}\\
\textbf{Renal Cell Carcinoma} & \\
786-O &  \cite{dhanabal1999, lieubeauteillet1998} \\
ACHN &  \cite{huang2008}\\
A-498 &  \cite{huang2008}\\
Caki-1 &  \cite{inoue2001, schirner1998}\\
&  \cite{shi2002}\\
SK-RC-29 &  \cite{prewett1998}\\
Caki-2 &  \cite{schirner1998}\\
Unspecified &  \cite{fujimoto1995}\\
\end{longtable} 


\section{Supplemental Materials: Results of Parameter Fittings}\label{App:3}


Individual data sets are labeled with the year and author, and given a unique identifier: either the label they were presented with in the figure from which the data originated, or the cell line that is used in the paper.

In some cases, the line representing the result of the parameter fitting is not visible. This happens for one of two reasons: a large difference between orders of magnitude in separate data sets, limiting the available space for data sets with smaller orders of magnitide; or because two or more data sets started with the same initial condition, causing the combined fitting result to produce the same curve. A complete list of all of the parameter fittings that are not visible, and the reason for why they cannot be seen, is given below:
\begin{itemize}
\item In the combined \textit{in vitro} bladder cancer trials, the AS clusterin and MM control trials of Miyake 2001 share an initial condition, hence only the MM control fitting is visible.
\item In the combined \textit{in vitro} breast cancer trials, the three Smith 2004 trials share the same initial condition, so the purple curve indicates the fitting to all three of these trials.
\item In the combined \textit{in vivo} breast cancer trials, the two Coopman 2000 trials share an initial condition, thus the green curve represents the fitting to both trials.
\item In the combined \textit{in vitro} colon cancer trials, the Moser and HCT116 trials have the same initial condition, so the green curve represents the combined fitting to both.
\item In the combined \textit{in vivo} colon cancer trials, two sets of trials have the same initial condition---the two Reinmuth 2002 trials and the two Warren 1995 trials. As a result, the orange curves account for both Reinmuth 2002 trials and the pink curves to both Warren 1995 trials.
\item In the combined \textit{in vivo} head and neck squamous cell carcinoma trials, the three Liu 1999 trials start with the same initial conditions, hence the teal curve represents the combined fitting to all three data sets.
\item In the combined \textit{in vitro} hepatocellular carcinoma trials, the Huynh 2008 trials share an initial condition, so the green curve represents the fitting to both data sets.
\item In the combined \textit{in vivo} hepatocellular carcinoma trials, the Liu 2005 data sets have the same initial condition, so the teal curve indicates the fitting to both data sets.
\item In the individual \textit{in vitro} lung cancer trials, Fig. 4D from Fabbri 2005 was cropped from the graph because it was two orders of magnitude higher than the next largest tumor, making the other 12 trials impossible to distinguish. Despite its exclusion here, it was used in the fitting analysis.
\item In the combined \textit{in vitro} lung cancer trials, not only is Fig. 4D from Fabbri 2005 excluded, but several trials from the same study have the same initial conditions (\emph{i.e.}, the four visible Fabbri 2005 trials, SW-900 and A549 from Esquela-Kerscher, and all three Fujiwara trials.) For this figure, the seafoam green curve is the fit for all 4 visible Fabbri 2005 trials, the SW-900 and A549 trials from Esquela-Kerscher are both represented by the yellow curve, and all three Fujiwara trials are represented by the purple curve. Additionally, for the von Bertalanffy fitting, the data sets from Fujiwara 1993 and Takahashi 1992 are hidden by Fig. 3 from Tsubouchi 2000, presumably because their initial conditions are sufficiently close to each other.
\item In the \textit{in vivo} melanoma trials, the Boucherke 1989 trial is difficult to see because of its relatively low order of magnitude, but is visible along the bottom of the graphs.
\item In the combined \textit{in vitro} ovarian cancer trials, the A2780 and SKOV-3 trials have the same initial condition, and the IGROV-1 and HCT-116 trials have the same initial condition. As a result, the green curve represents the fitting to the first two trials, and the purple curve is the fitting to the last two trials.
\end{itemize}

\begin{landscape}
\begin{figure}
    \begin{center}
    \hspace{-4cm}
      \includegraphics[width=1.1\linewidth]{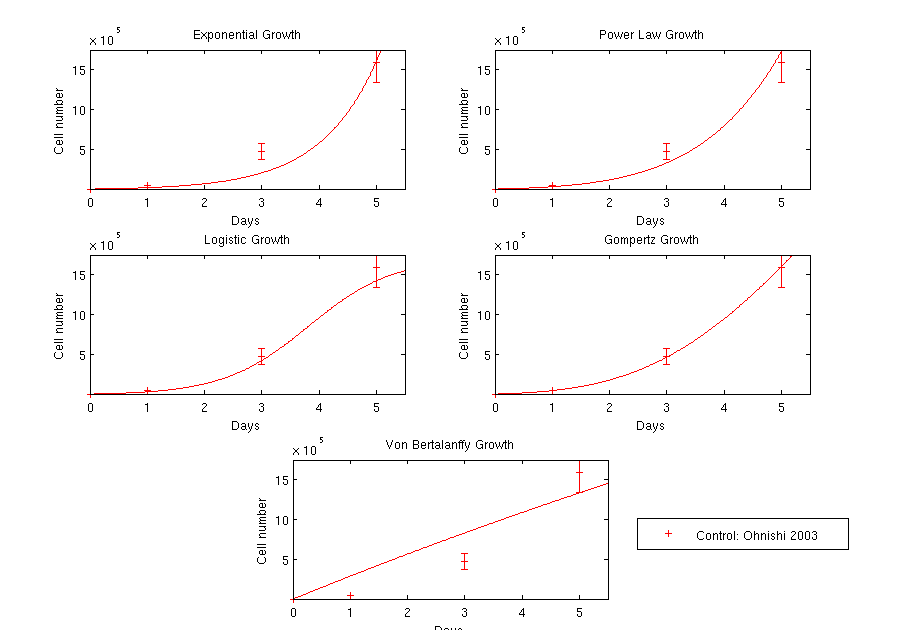}
    \end{center}
    \caption{Parameter Fittings to \textit{In Vitro} Bladder
      Cancer Trials}
      \end{figure}
\end{landscape}

\begin{landscape}
\begin{figure}
    \begin{center}
    \hspace{-4cm}
      \includegraphics[width=1.1\linewidth]{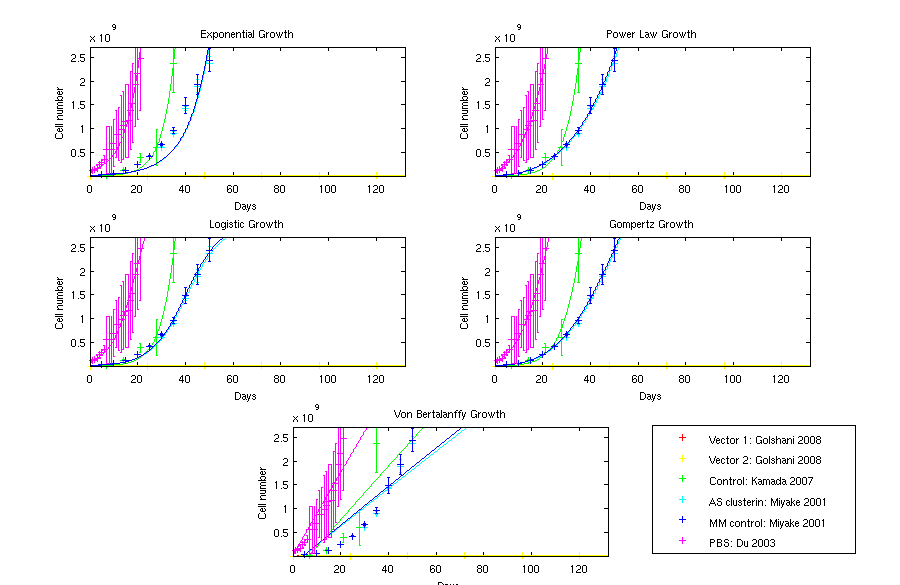}
    \end{center}
    \caption{Parameter Fittings to Individual \textit{In Vivo}
      Bladder Cancer Trials}
      \end{figure}
\end{landscape}

\begin{landscape}
\begin{figure}
    \begin{center}
    \hspace{-4cm}
      \includegraphics[width=1.1\linewidth]{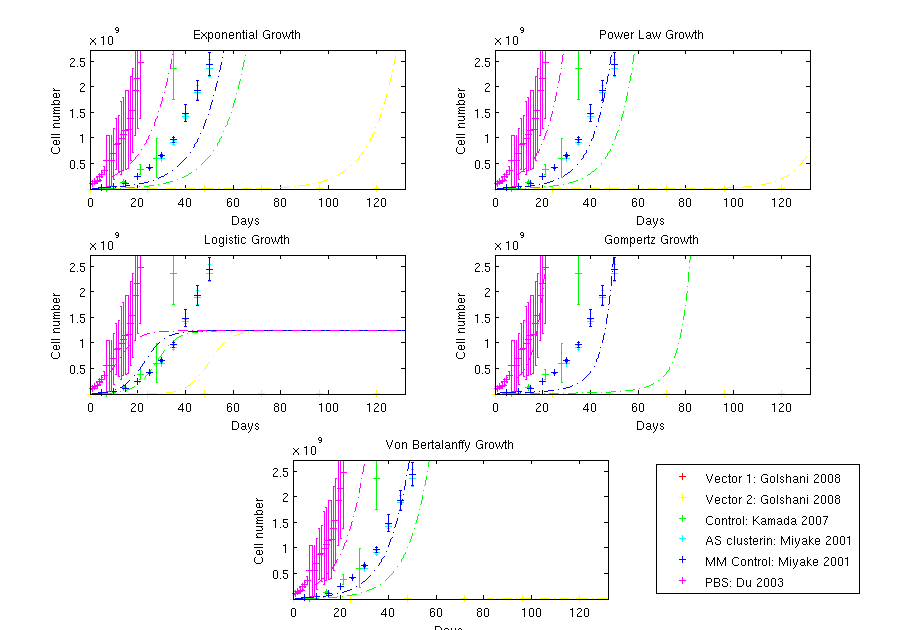}
    \end{center}
    \caption{Parameter Fitting to Combined \textit{In Vivo}
      Bladder Cancer Trials}
      \end{figure}
\end{landscape}

\begin{landscape}
\begin{figure}
\begin{center}
\hspace{-4cm}
\includegraphics[width=1.1\linewidth]{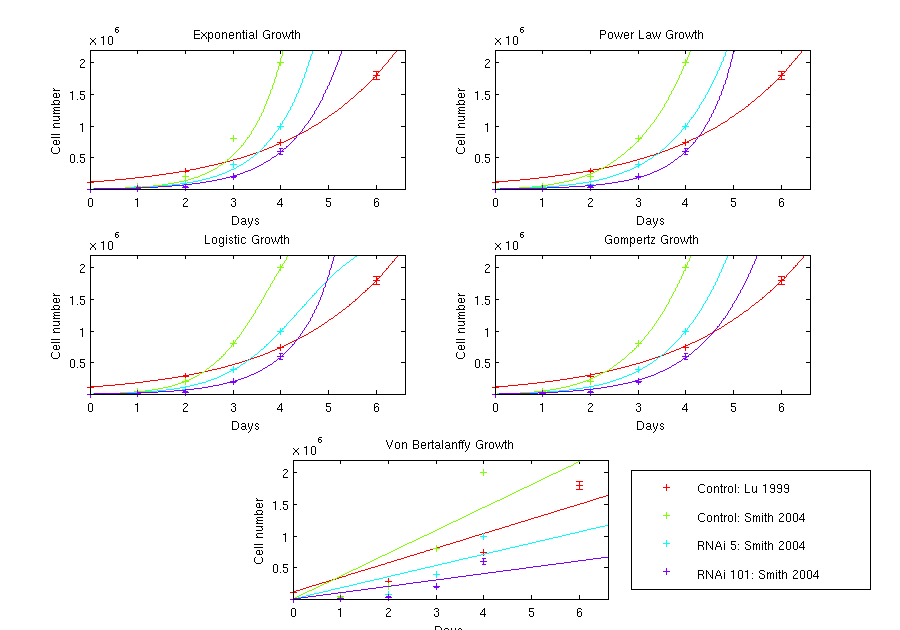}
\end{center}
\caption{Parameter Fittings to Individual \textit{In Vitro} Breast Cancer Trials}
\end{figure}
\end{landscape}

\begin{landscape}
\begin{figure}
\begin{center}
\hspace{-4cm}
\includegraphics[width=1.1\linewidth]{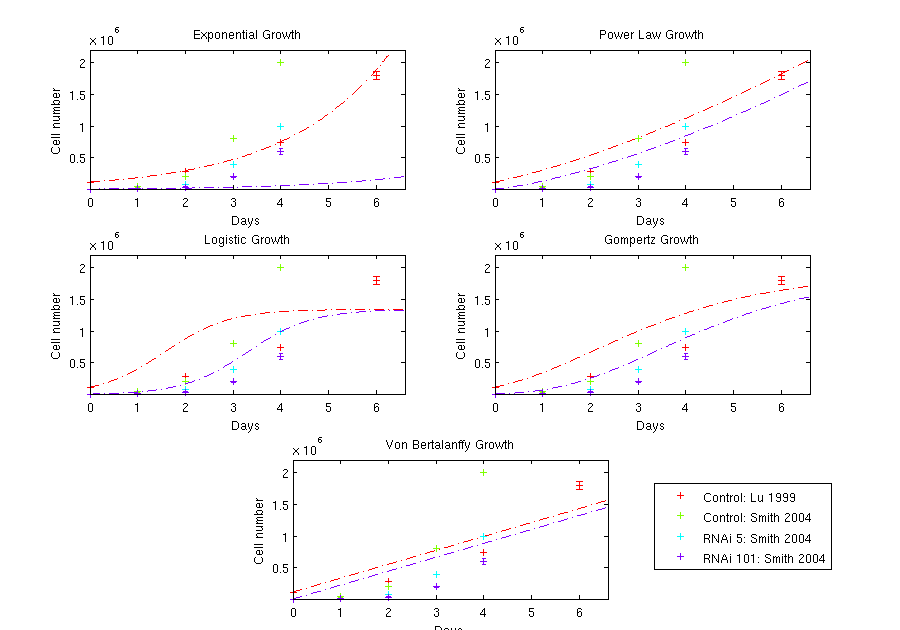}
\end{center}
\caption{Parameter Fitting to Combined \textit{In Vitro} Breast Cancer Trials}
\end{figure}
\end{landscape}

\begin{landscape}
\begin{figure}
\begin{center}
\hspace{-4cm}
\includegraphics[width=1.1\linewidth]{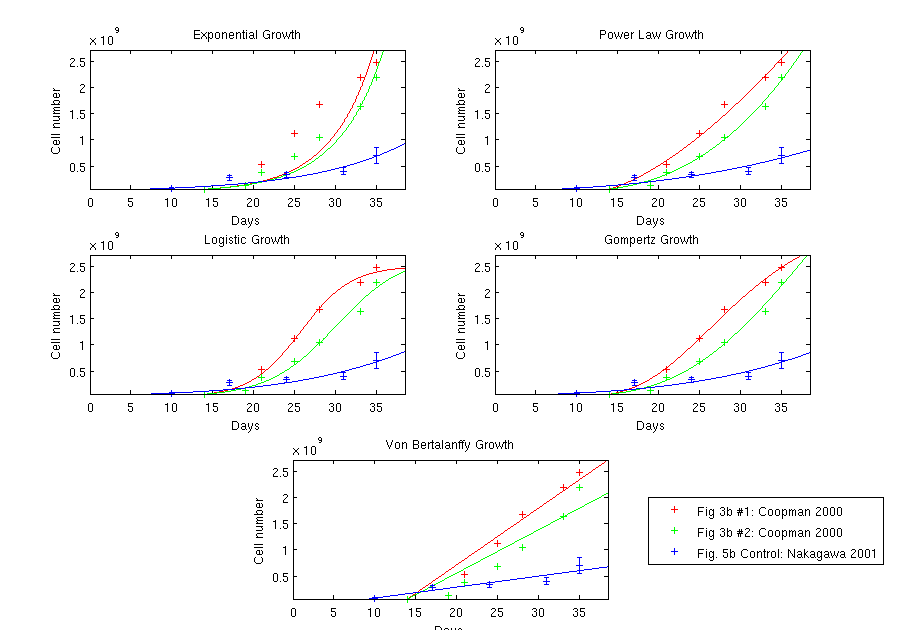}
\end{center}
\caption{Parameter Fittings to Individual \textit{In Vivo} Breast Cancer Trials}
\end{figure}
\end{landscape}

\begin{landscape}
\begin{figure}
\begin{center}
\hspace{-4cm}
\includegraphics[width=1.1\linewidth]{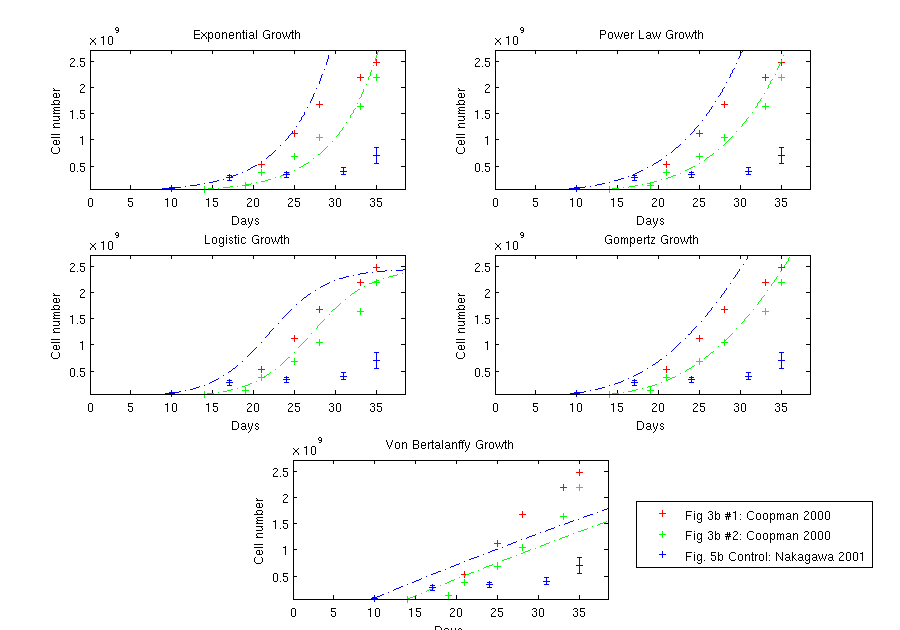}
\end{center}
\caption{Parameter Fitting to Combined \textit{In Vivo} Breast Cancer Trials}
\end{figure}
\end{landscape}

\begin{landscape}
\begin{figure}
\begin{center}
\hspace{-4cm}
\includegraphics[width=1.1\linewidth]{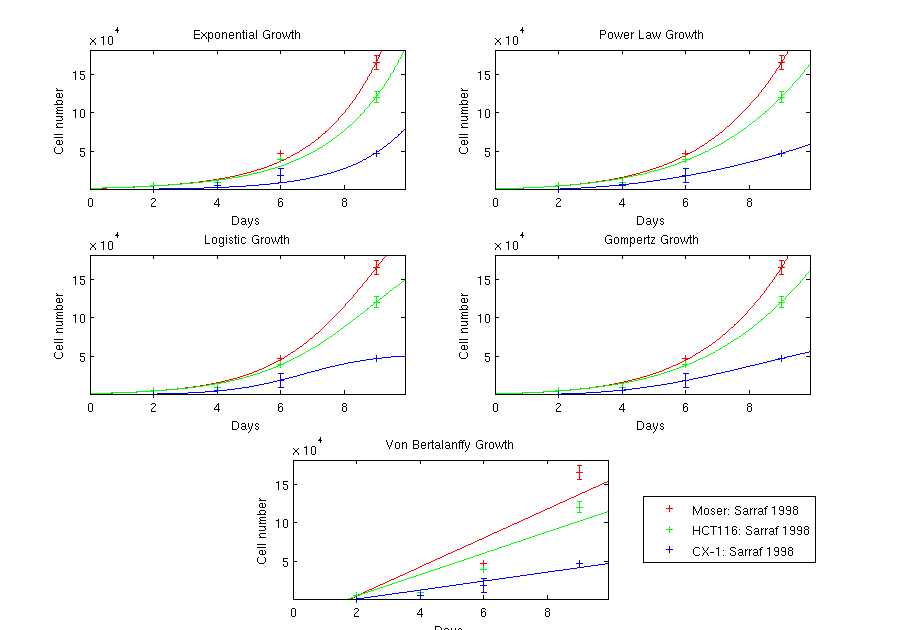}
\end{center}
\caption{Parameter Fittings to Individual \textit{In Vitro} Colon Cancer Trials}
\end{figure}
\end{landscape}

\begin{landscape}
\begin{figure}
\begin{center}
\hspace{-4cm}
\includegraphics[width=1.1\linewidth]{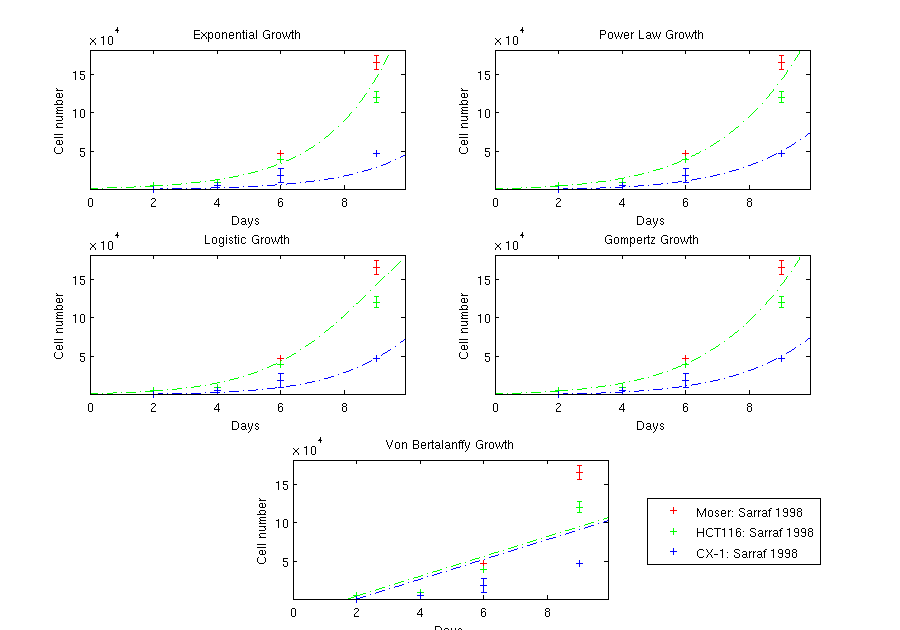}
\end{center}
\caption{Parameter Fitting to Combined \textit{In Vitro} Colon Cancer Trials}
\end{figure}
\end{landscape}

\begin{landscape}
\begin{figure}
\begin{center}
\hspace{-4cm}
\includegraphics[width=1.1\linewidth]{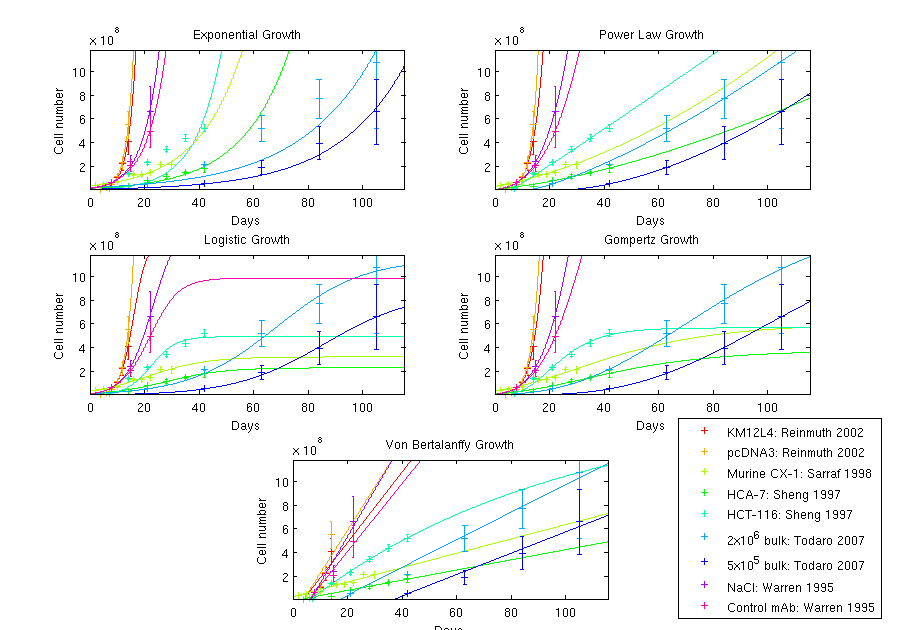}
\end{center}
\caption{Parameter Fittings to Individual \textit{In Vivo} Colon Cancer Trials}
\end{figure}
\end{landscape}

\begin{landscape}
\begin{figure}
\begin{center}
\hspace{-4cm}
\includegraphics[width=1.1\linewidth]{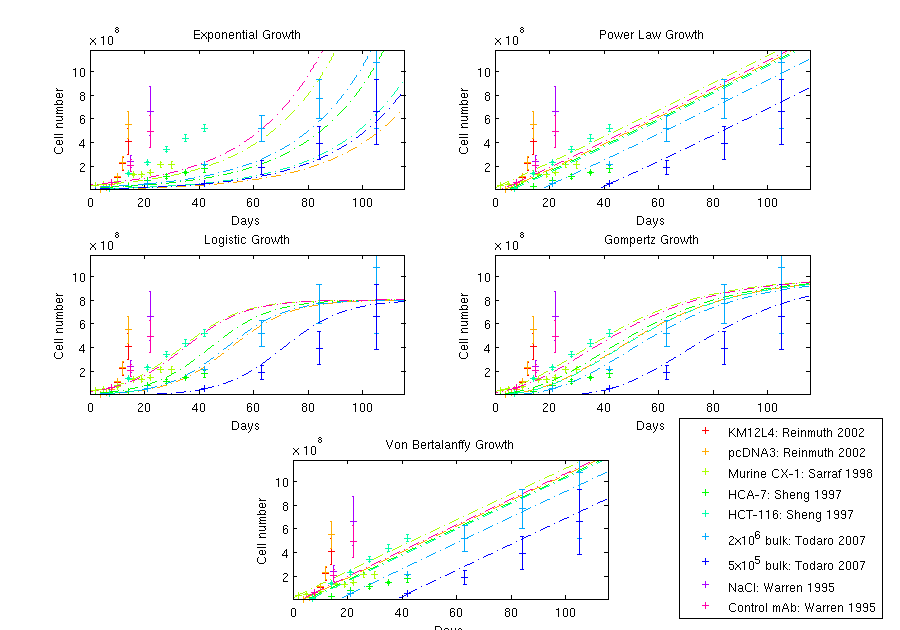}
\end{center}
\caption{Parameter Fitting to Combined \textit{In Vivo} Colon Cancer Trials}
\end{figure}
\end{landscape}

\begin{landscape}
\begin{figure}
\begin{center}
\hspace{-4cm}
\includegraphics[width=1.1\linewidth]{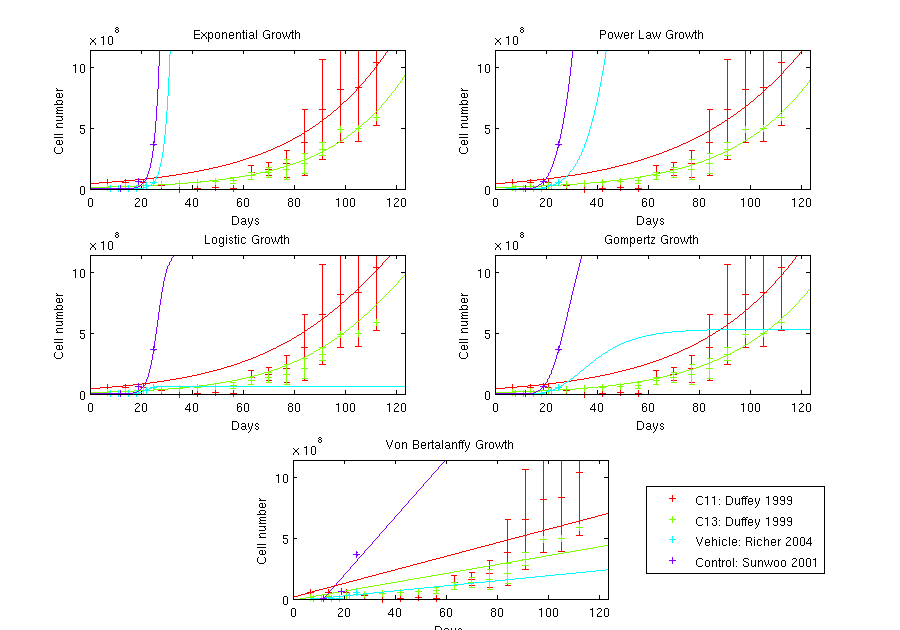}
\end{center}
\caption{Parameter Fittings to Individual \textit{In Vitro} Head and Neck Squamous Cell Carcinoma Trials}
\end{figure}
\end{landscape}

\begin{landscape}
\begin{figure}
\begin{center}
\hspace{-4cm}
\includegraphics[width=1.1\linewidth]{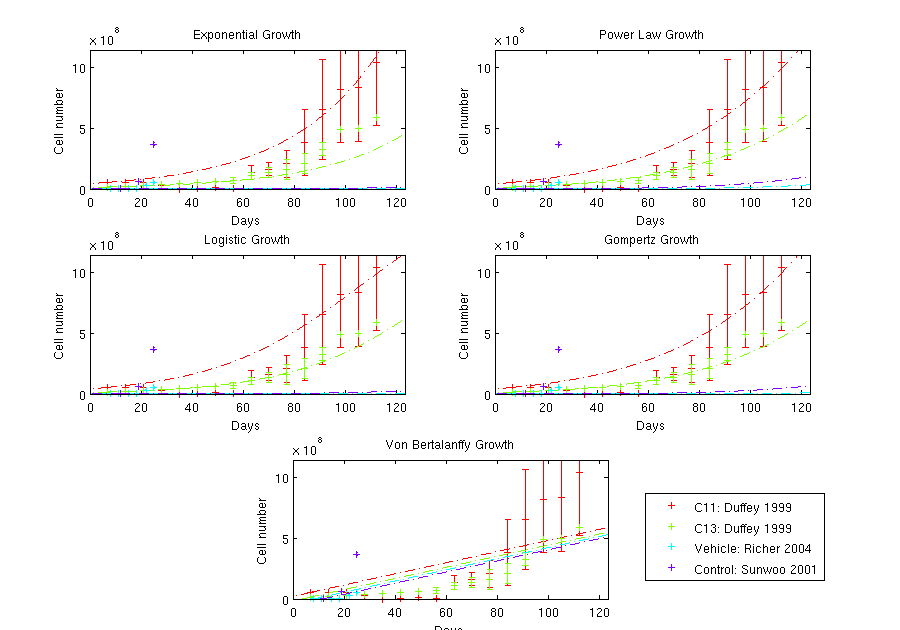}
\end{center}
\caption{Parameter Fitting to Combined \textit{In Vitro} Head and Neck Squamous Cell Carcinoma Trials}
\end{figure}
\end{landscape}

\begin{landscape}
\begin{figure}
\begin{center}
\hspace{-4cm}
\includegraphics[width=1.1\linewidth]{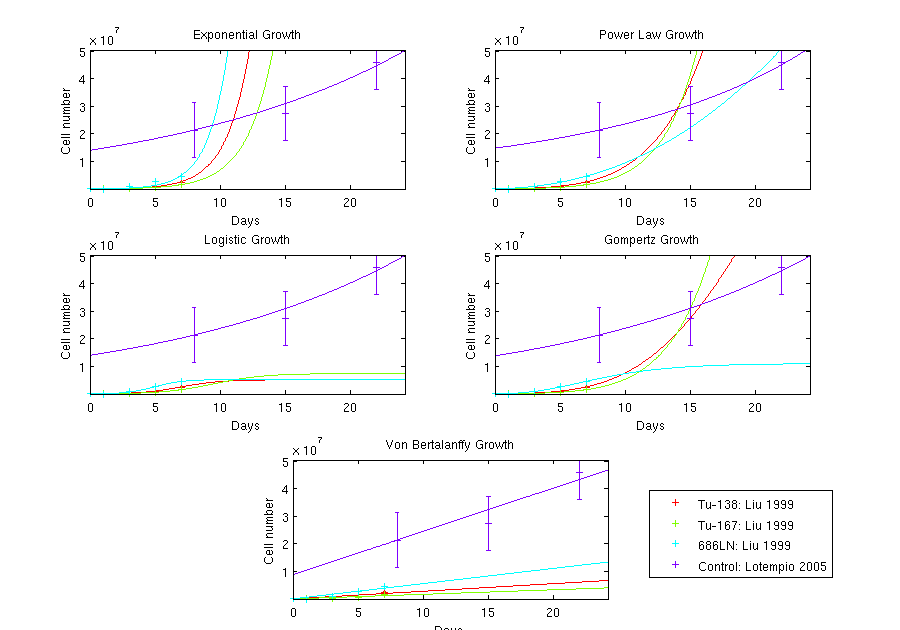}
\end{center}
\caption{Parameter Fittings to Individual \textit{In Vivo} Head and Neck Squamous Cell Carcinoma Trials}
\end{figure}
\end{landscape}

\begin{landscape}
\begin{figure}
\begin{center}
\hspace{-4cm}
\includegraphics[width=1.1\linewidth]{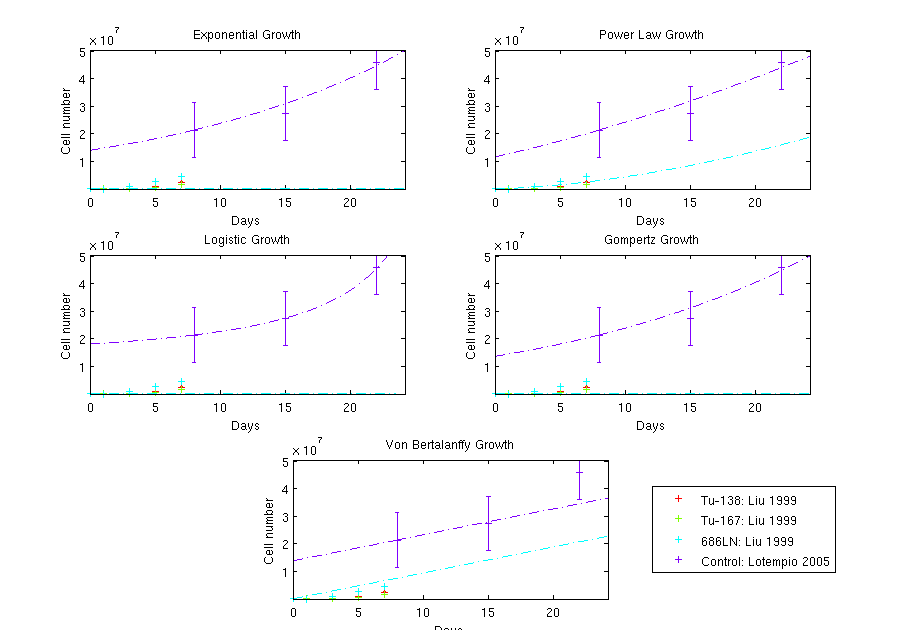}
\end{center}
\caption{Parameter Fitting to Combined \textit{In Vivo} Head and Neck Squamous Cell Carcinoma Trials}
\end{figure}
\end{landscape}

\begin{landscape}
\begin{figure}
\begin{center}
\hspace{-4cm}
\includegraphics[width=1.1\linewidth]{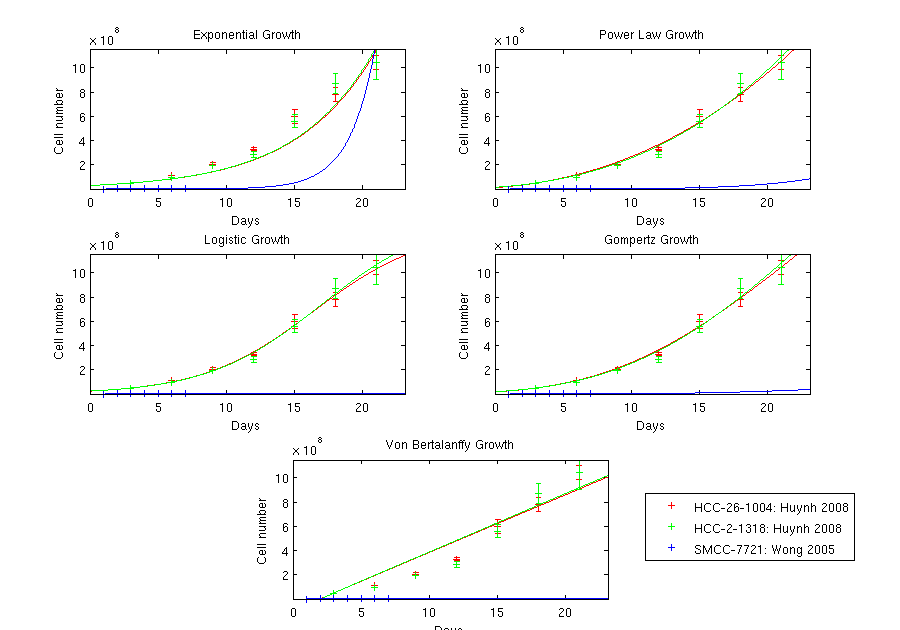}
\end{center}
\caption{Parameter Fittings to Individual \textit{In Vitro} Hepatocellular Carcinoma Trials}
\end{figure}
\end{landscape}

\begin{landscape}
\begin{figure}
\begin{center}
\hspace{-4cm}
\includegraphics[width=1.1\linewidth]{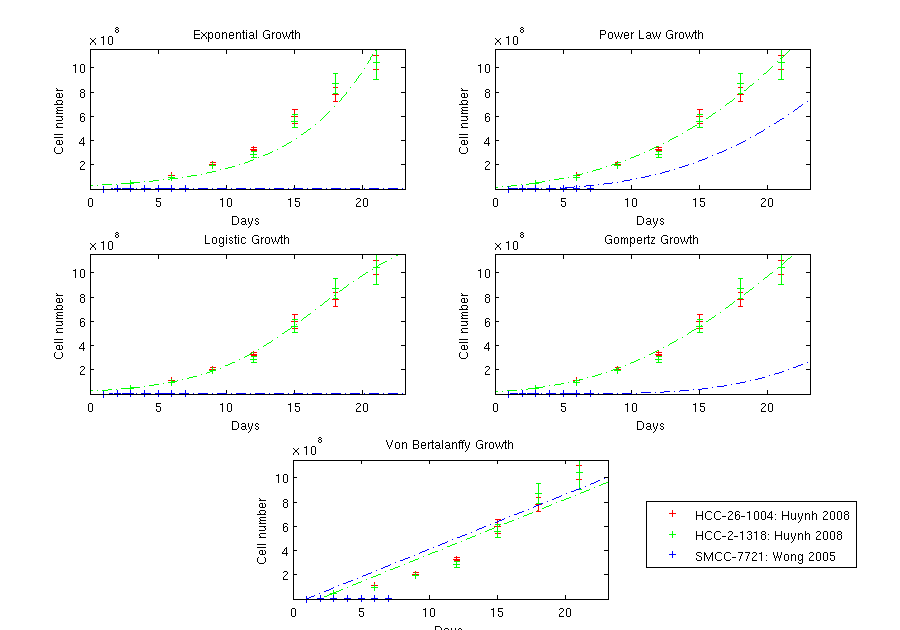}
\end{center}
\caption{Parameter Fitting to Combined \textit{In Vitro} Hepatocellular Carcinoma Trials}
\end{figure}
\end{landscape}

\begin{landscape}
\begin{figure}
\begin{center}
\hspace{-4cm}
\includegraphics[width=1.1\linewidth]{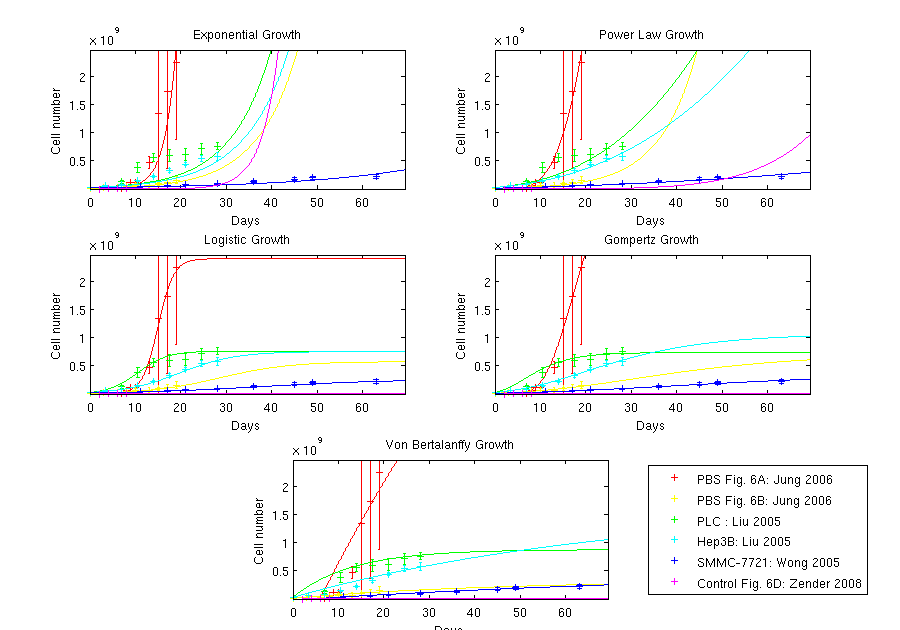}
\end{center}
\caption{Parameter Fittings to Individual \textit{In Vivo} Hepatocellular Carcinoma Trials}
\end{figure}
\end{landscape}

\begin{landscape}
\begin{figure}
\begin{center}
\hspace{-4cm}
\includegraphics[width=1.1\linewidth]{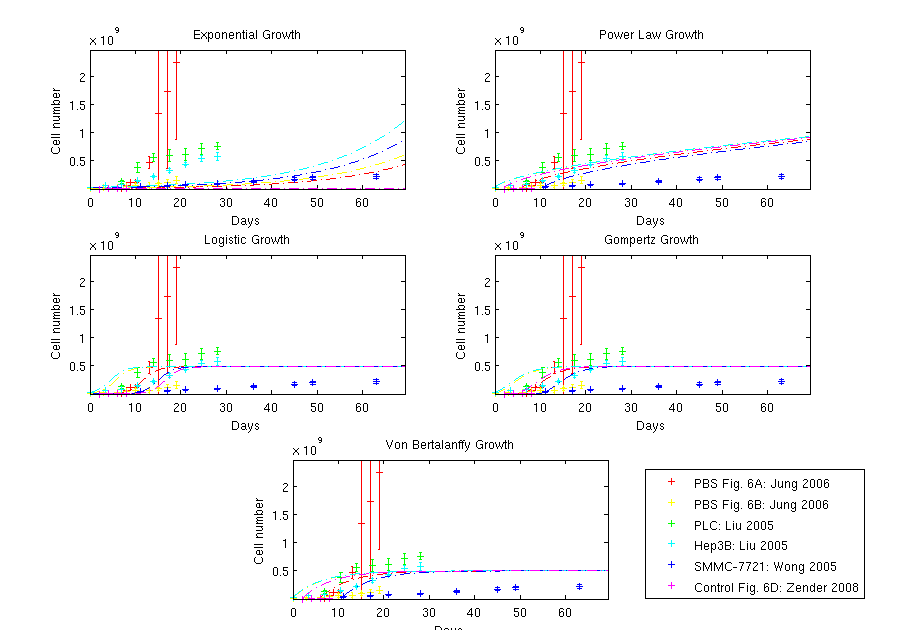}
\end{center}
\caption{Parameter Fitting to Combined \textit{In Vivo} Hepatocellular Carcinoma Trials}
\end{figure}
\end{landscape}

\begin{landscape}
\begin{figure}
\begin{center}
\hspace{-4cm}
\includegraphics[width=1.1\linewidth]{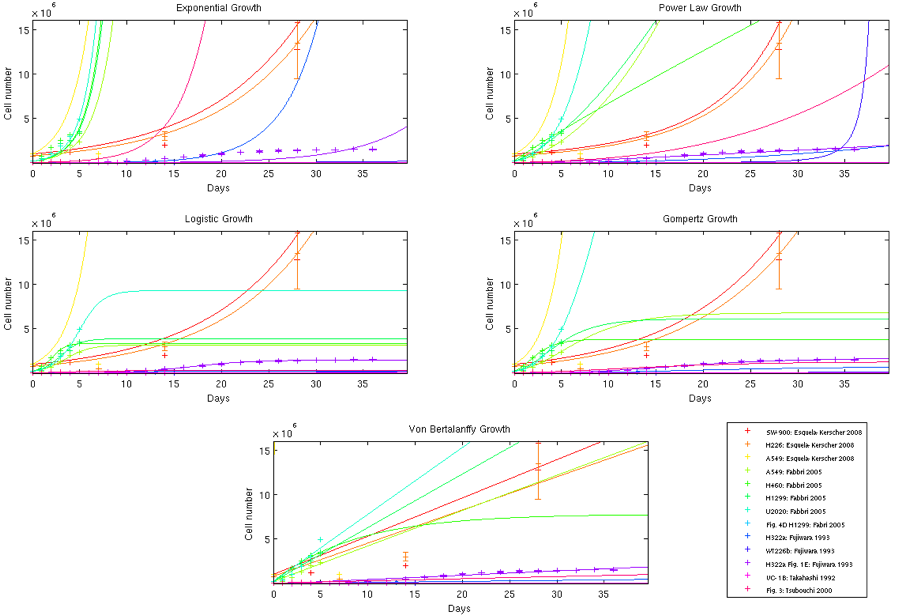}
\end{center}
\caption{Parameter Fittings to Individual \textit{In Vitro} Lung Cancer Trials}
\end{figure}
\end{landscape}

\begin{landscape}
\begin{figure}
\begin{center}
\hspace{-4cm}
\includegraphics[width=1.1\linewidth]{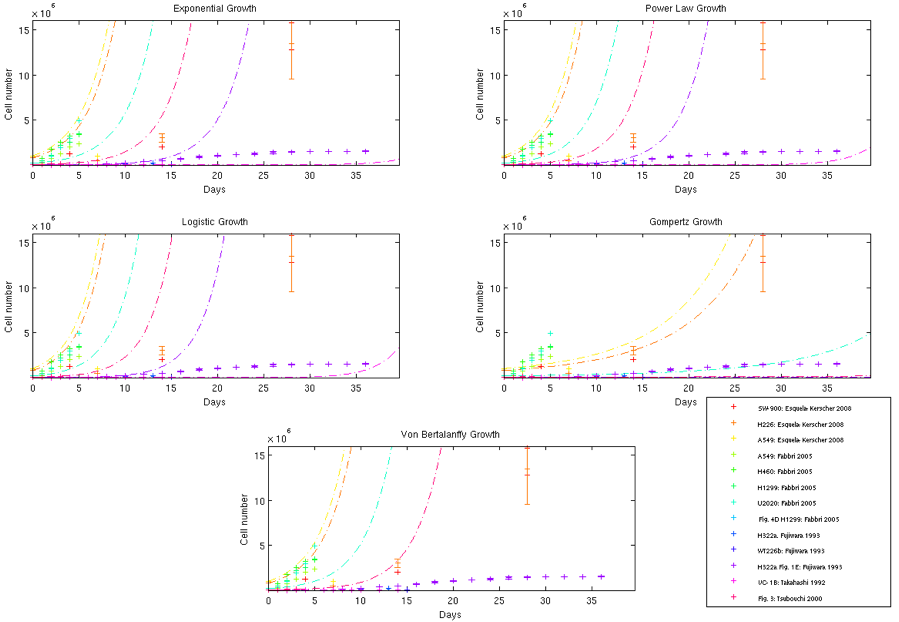}
\end{center}
\caption{Parameter Fitting to Combined \textit{In Vitro} Lung Cancer Trials}
\end{figure}
\end{landscape}

\begin{landscape}
\begin{figure}
\begin{center}
\hspace{-4cm}
\includegraphics[width=1.1\linewidth]{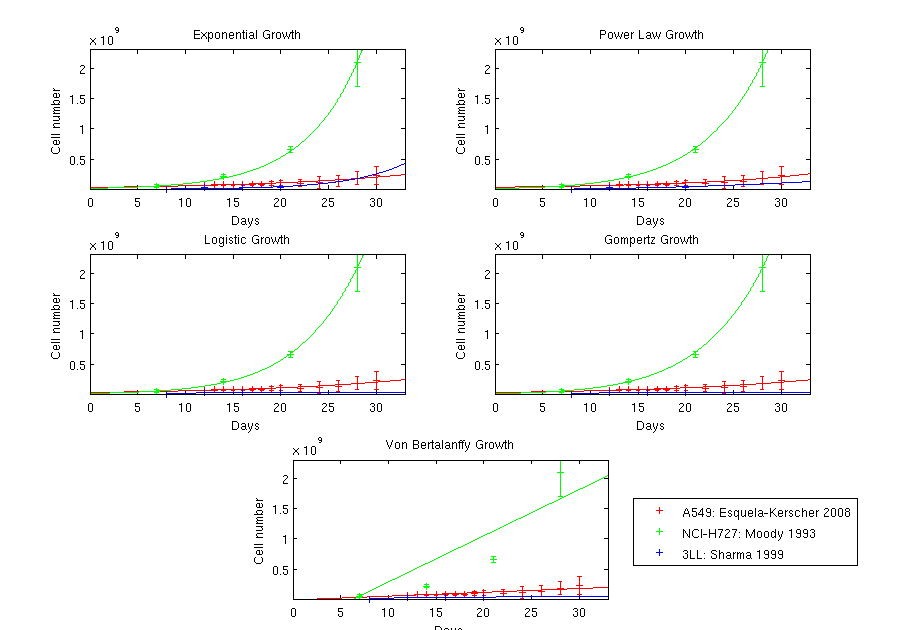}
\end{center}
\caption{Parameter Fittings to Individual \textit{In Vivo} Lung Cancer Trials}
\end{figure}
\end{landscape}

\begin{landscape}
\begin{figure}
\begin{center}
\hspace{-4cm}
\includegraphics[width=1.1\linewidth]{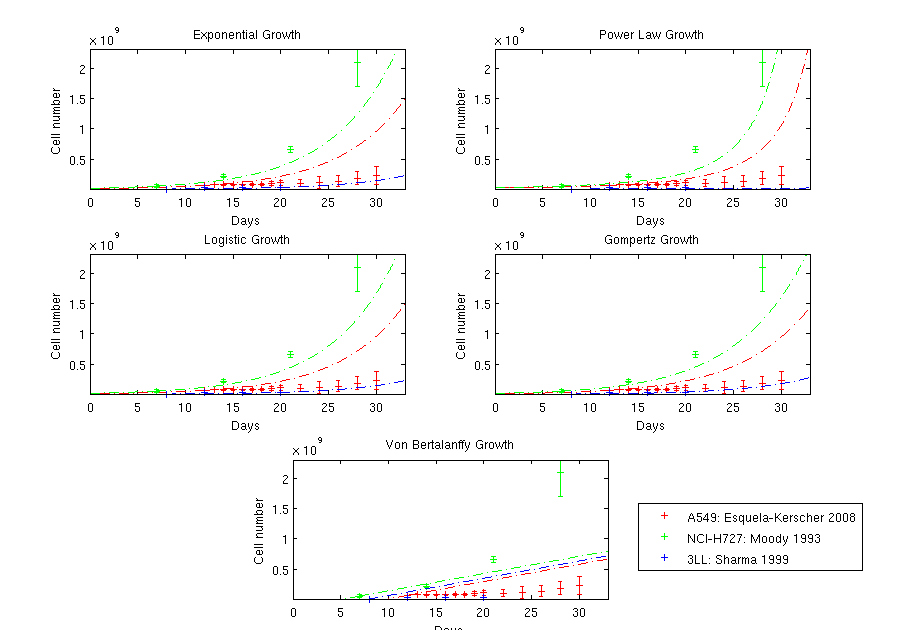}
\end{center}
\caption{Parameter Fitting to Combined \textit{In Vivo} Lung Cancer Trials}
\end{figure}
\end{landscape}

\begin{landscape}
\begin{figure}
\begin{center}
\hspace{-4cm}
\includegraphics[width=1.1\linewidth]{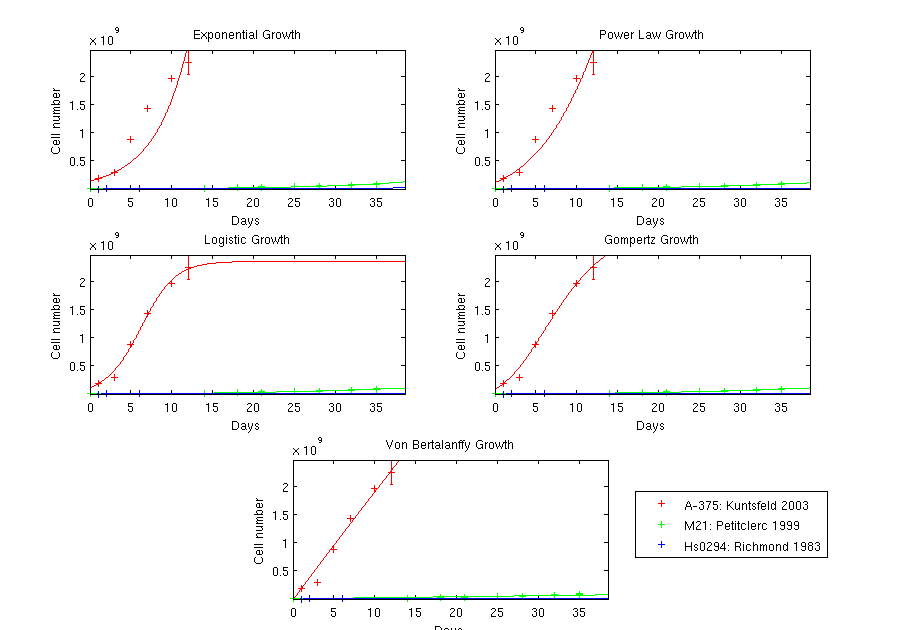}
\end{center}
\caption{Parameter Fittings to Individual \textit{In Vitro} Melanoma Trials}
\end{figure}
\end{landscape}

\begin{landscape}
\begin{figure}
\begin{center}
\hspace{-4cm}
\includegraphics[width=1.1\linewidth]{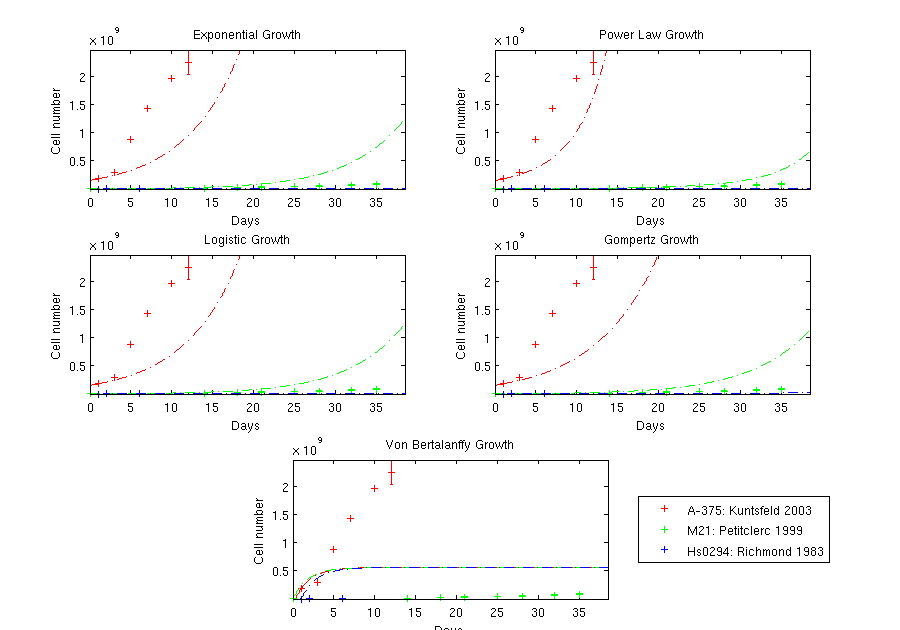}
\end{center}
\caption{Parameter Fitting to Combined \textit{In Vitro} Melanoma Trials}
\end{figure}
\end{landscape}

\begin{landscape}
\begin{figure}
\begin{center}
\hspace{-4cm}
\includegraphics[width=1.1\linewidth]{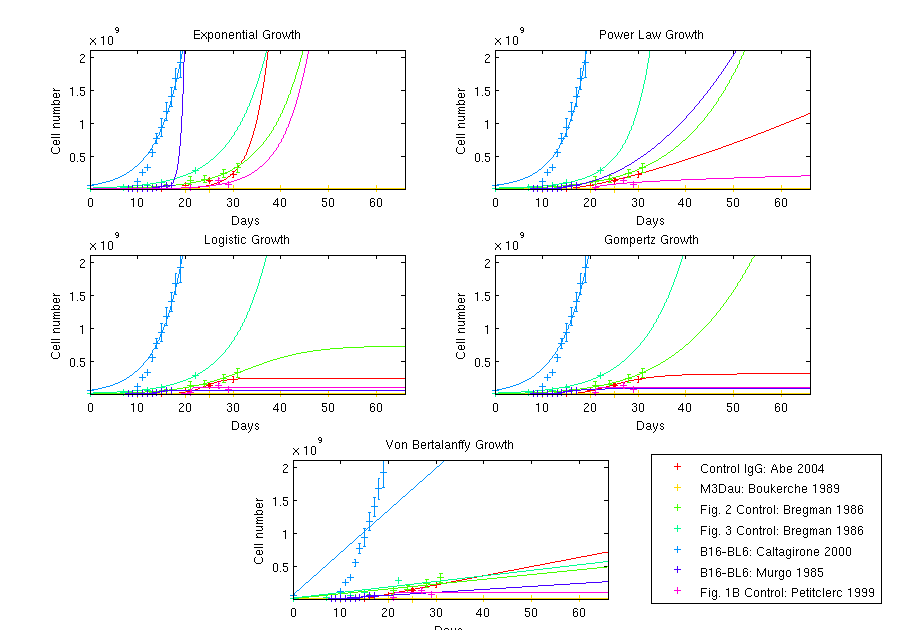}
\end{center}
\caption{Parameter Fittings to Individual \textit{In Vivo} Melanoma Trials}
\end{figure}
\end{landscape}

\begin{landscape}
\begin{figure}
\begin{center}
\hspace{-4cm}
\includegraphics[width=1.1\linewidth]{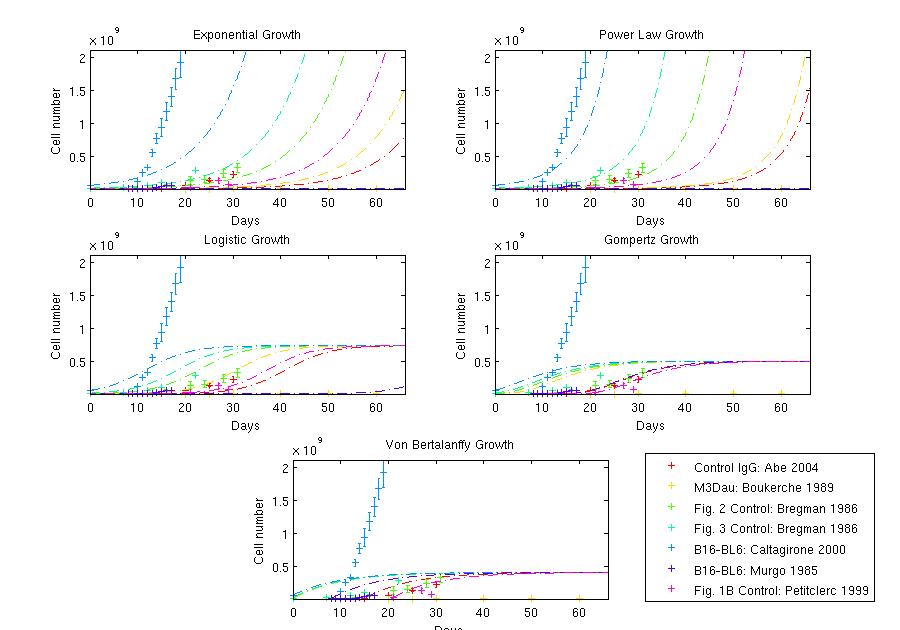}
\end{center}
\caption{Parameter Fitting to Combined \textit{In Vivo} Melanoma Trials}
\end{figure}
\end{landscape}

\begin{landscape}
\begin{figure}
\begin{center}
\hspace{-4cm}
\includegraphics[width=1.1\linewidth]{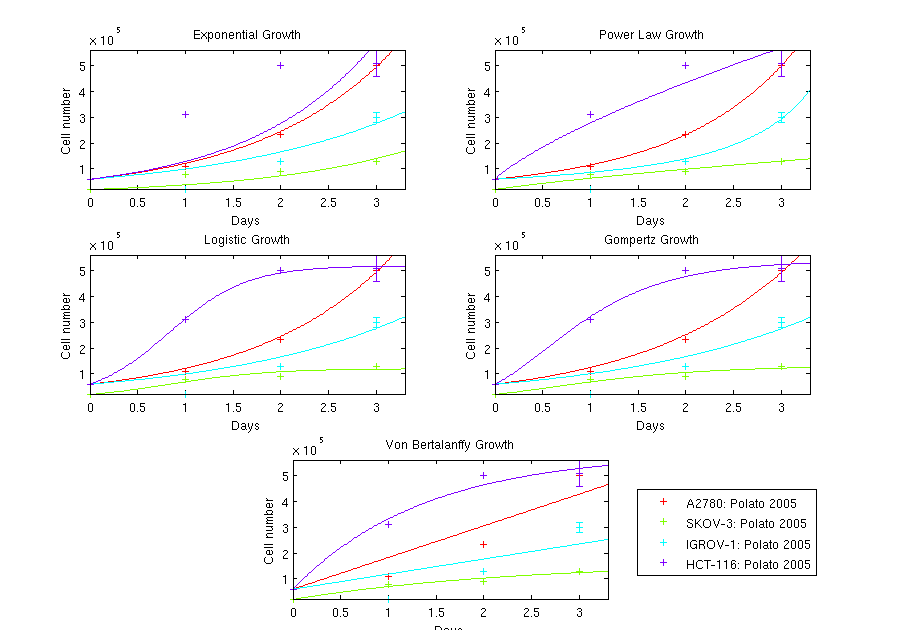}
\end{center}
\caption{Parameter Fittings to Individual \textit{In Vitro} Ovarian Cancer Trials}
\end{figure}
\end{landscape}

\begin{landscape}
\begin{figure}
\begin{center}
\hspace{-4cm}
\includegraphics[width=1.1\linewidth]{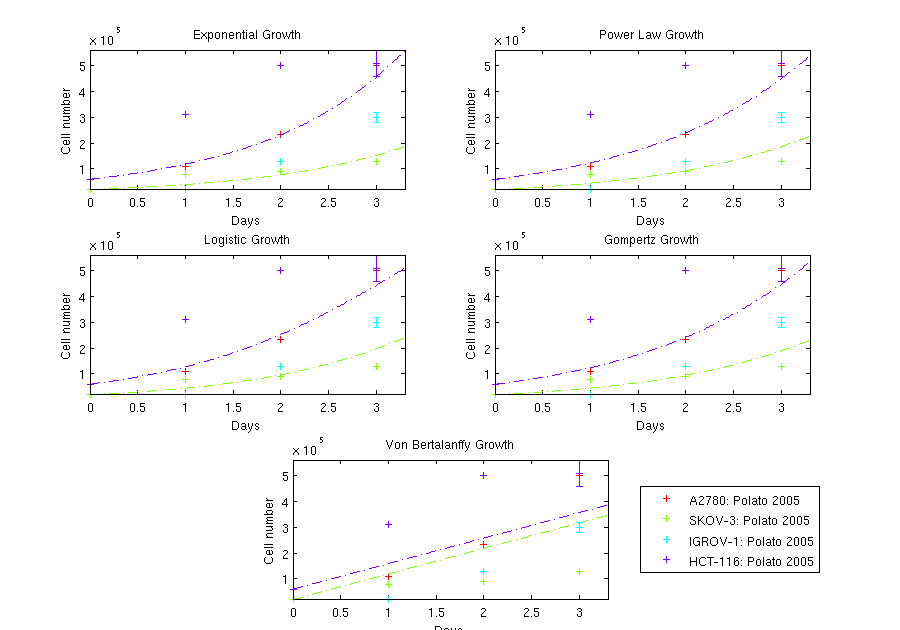}
\end{center}
\caption{Parameter Fitting to Combined \textit{In Vitro} Ovarian Cancer Trials}
\end{figure}
\end{landscape}

\begin{landscape}
\begin{figure}
\begin{center}
\hspace{-4cm}
\includegraphics[width=1.1\linewidth]{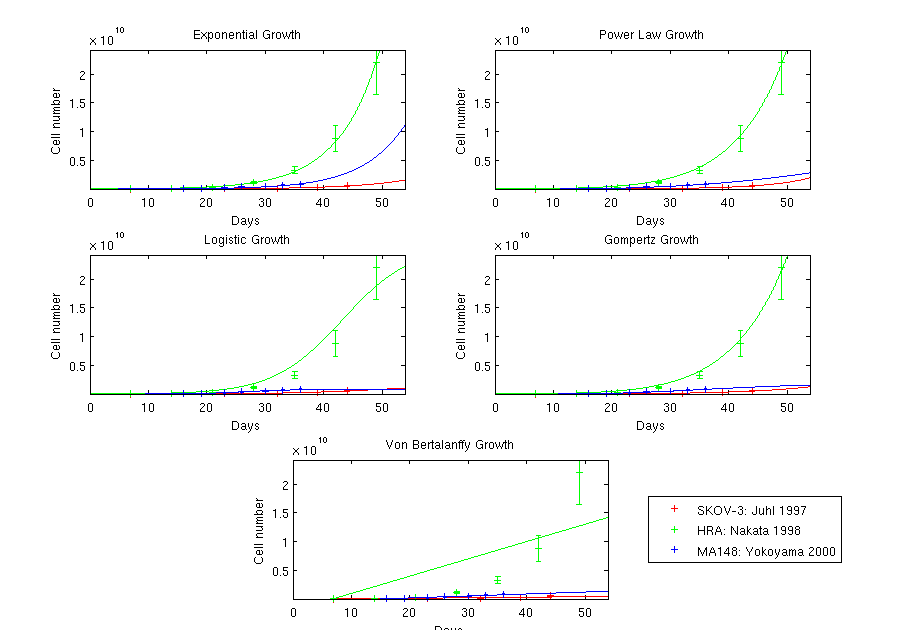}
\end{center}
\caption{Parameter Fittings to Individual \textit{In Vivo} Ovarian Cancer Trials}
\end{figure}
\end{landscape}

\begin{landscape}
\begin{figure}
\begin{center}
\hspace{-4cm}
\includegraphics[width=1.1\linewidth]{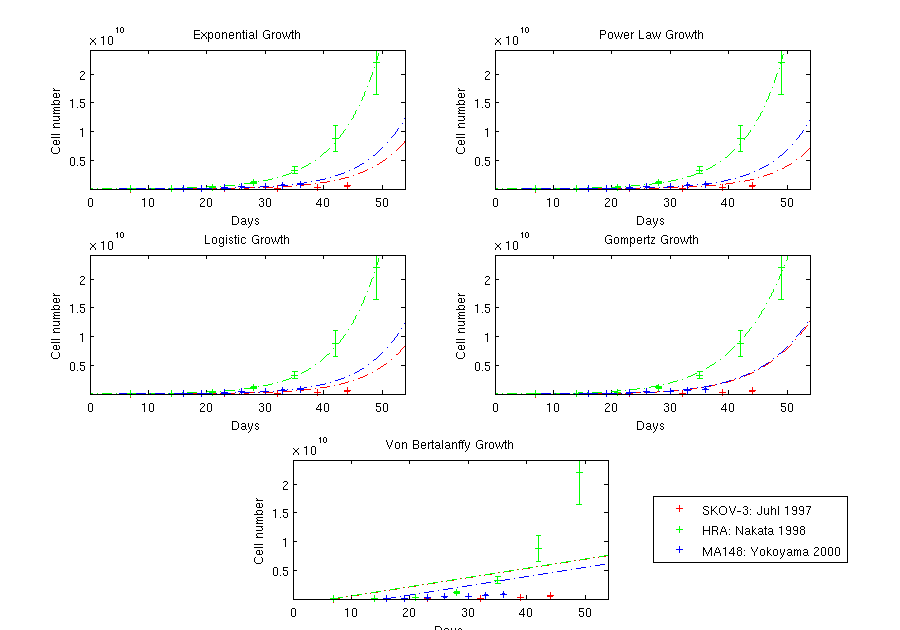}
\end{center}
\caption{Parameter Fitting to Combined \textit{In Vivo} Ovarian Cancer Trials}
\end{figure}
\end{landscape}

\begin{landscape}
\begin{figure}
\begin{center}
\hspace{-4cm}
\includegraphics[width=1.1\linewidth]{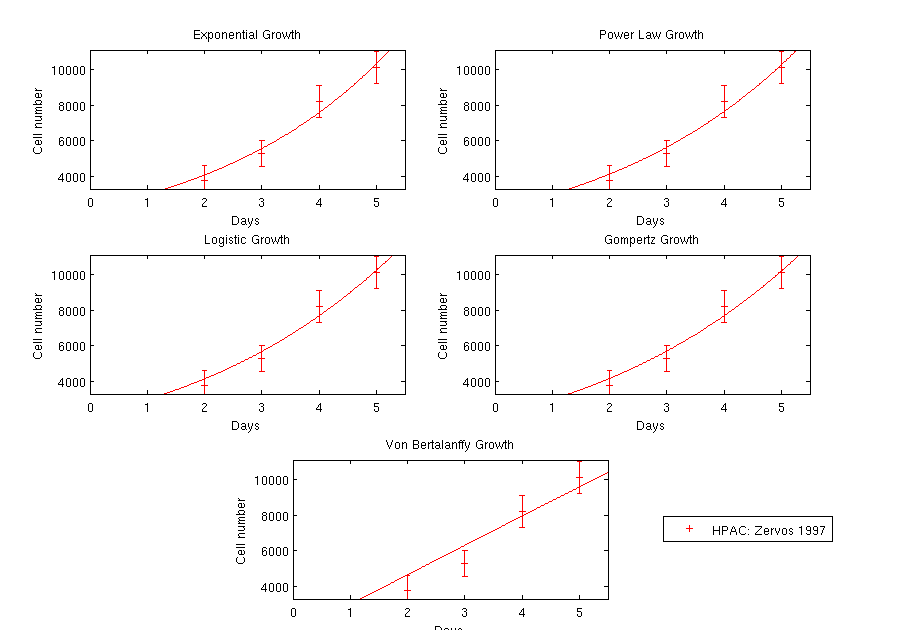}
\end{center}
\caption{Parameter Fittings to \textit{In Vitro} Pancreatic Cancer Trials}
\end{figure}
\end{landscape}

\begin{landscape}
\begin{figure}
\begin{center}
\hspace{-4cm}
\includegraphics[width=1.1\linewidth]{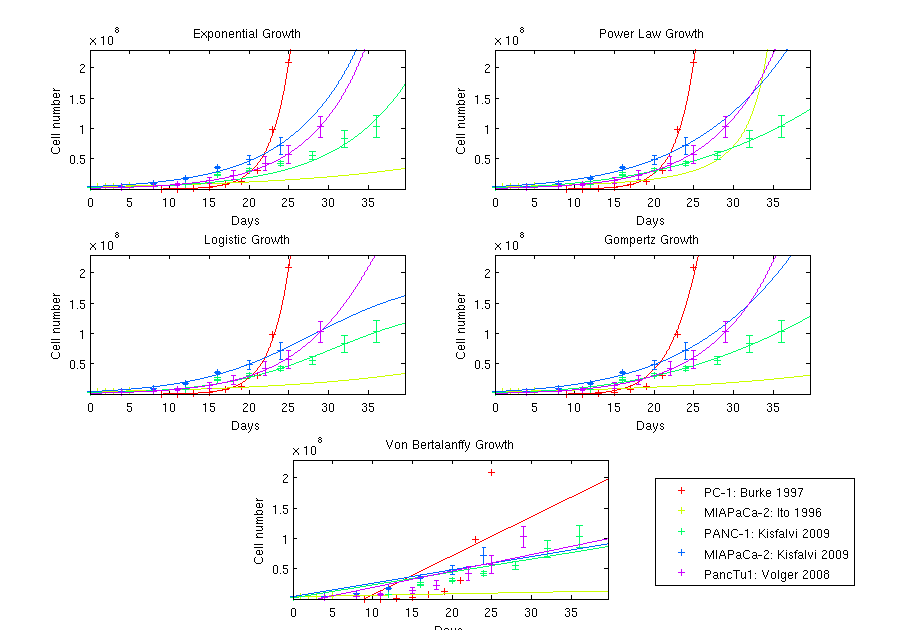}
\end{center}
\caption{Parameter Fittings to Individual \textit{In Vivo} Pancreatic Cancer Trials}
\end{figure}
\end{landscape}

\begin{landscape}
\begin{figure}
\begin{center}
\hspace{-4cm}
\includegraphics[width=1.1\linewidth]{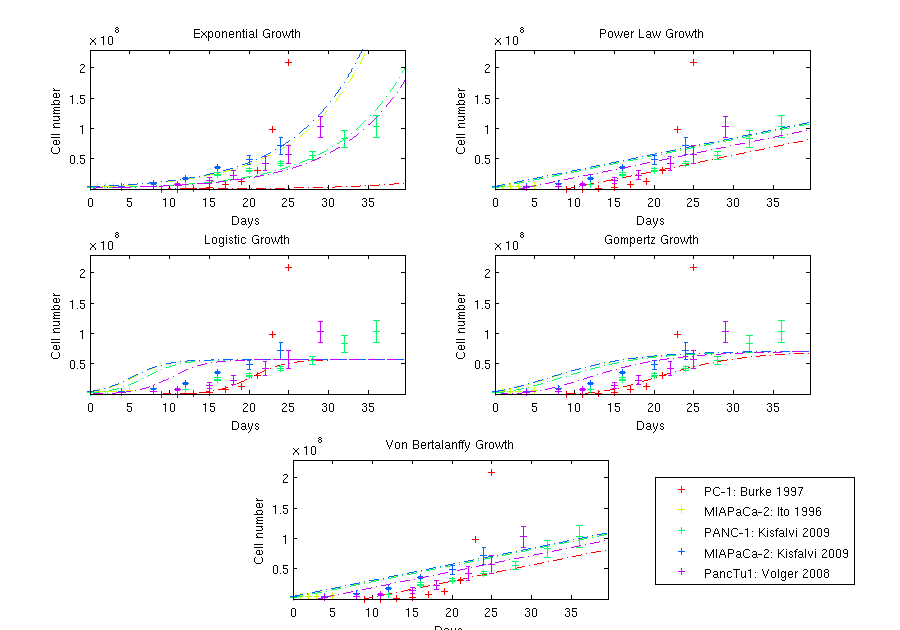}
\end{center}
\caption{Parameter Fitting to Combined \textit{In Vivo} Pancreatic Cancer Trials}
\end{figure}
\end{landscape}

\begin{landscape}
\begin{figure}
\begin{center}
\hspace{-4cm}
\includegraphics[width=1.1\linewidth]{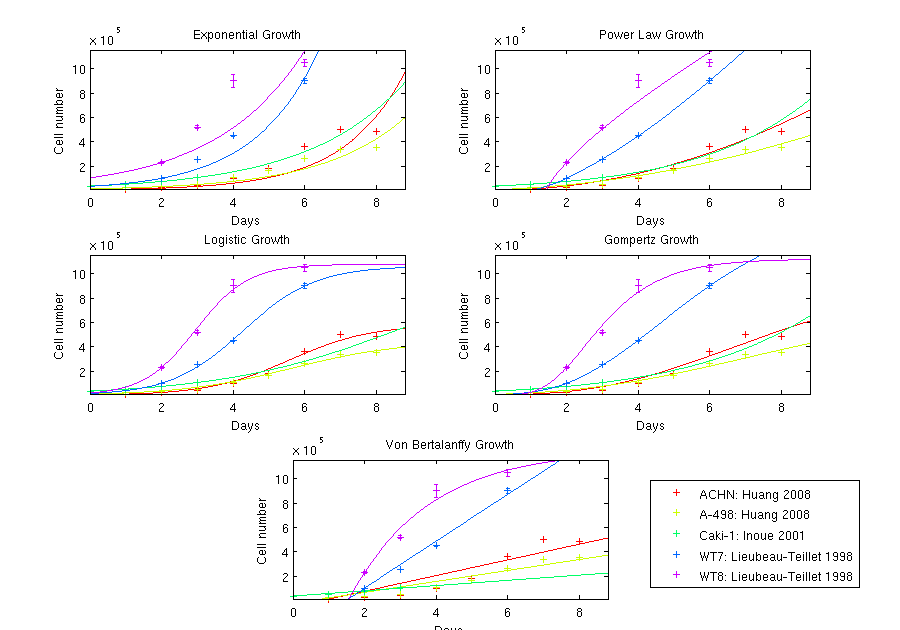}
\end{center}
\caption{Parameter Fittings to Individual \textit{In Vitro} Renal Cell Carcinoma Trials}
\end{figure}
\end{landscape}

\begin{landscape}
\begin{figure}
\begin{center}
\hspace{-4cm}
\includegraphics[width=1.1\linewidth]{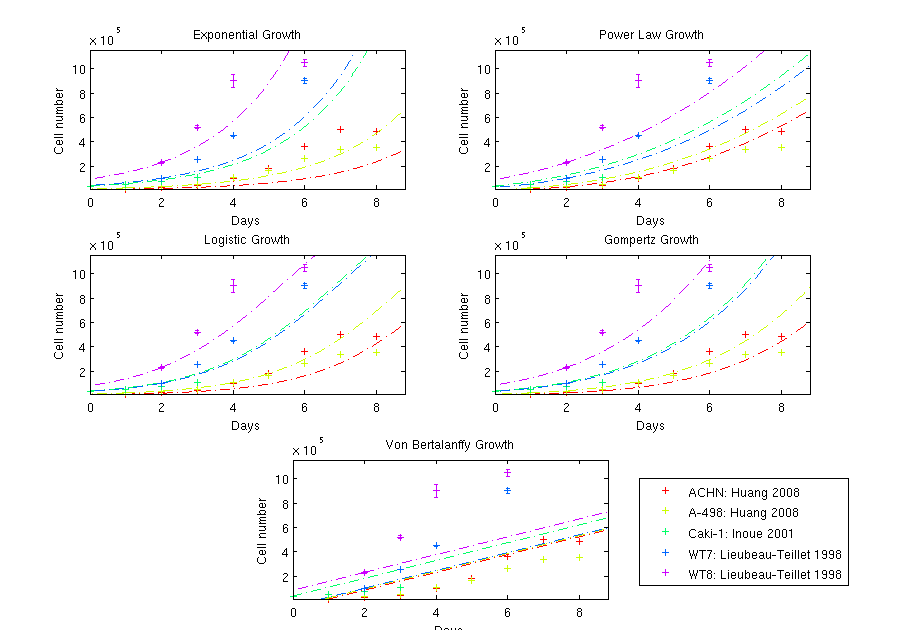}
\end{center}
\caption{Parameter Fitting to Combined \textit{In Vitro} Renal Cell Carcinoma Trials}
\end{figure}
\end{landscape}

\begin{landscape}
\begin{figure}
\begin{center}
\hspace{-4cm}
\includegraphics[width=1.1\linewidth]{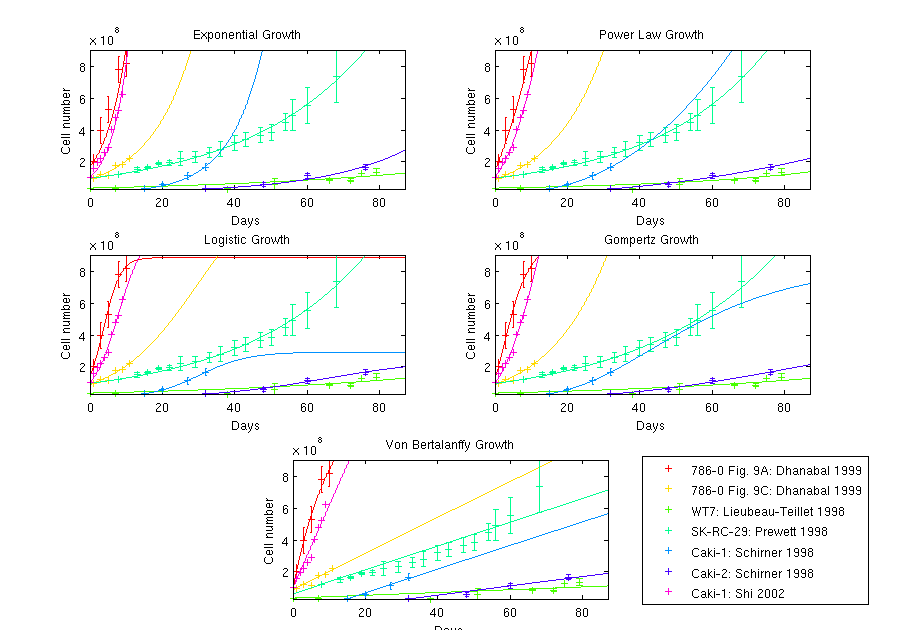}
\end{center}
\caption{Parameter Fittings to Individual \textit{In Vivo} Renal Cell Carcinoma Trials}
\end{figure}
\end{landscape}

\begin{landscape}
\begin{figure}
\begin{center}
\hspace{-4cm}
\includegraphics[width=1.1\linewidth]{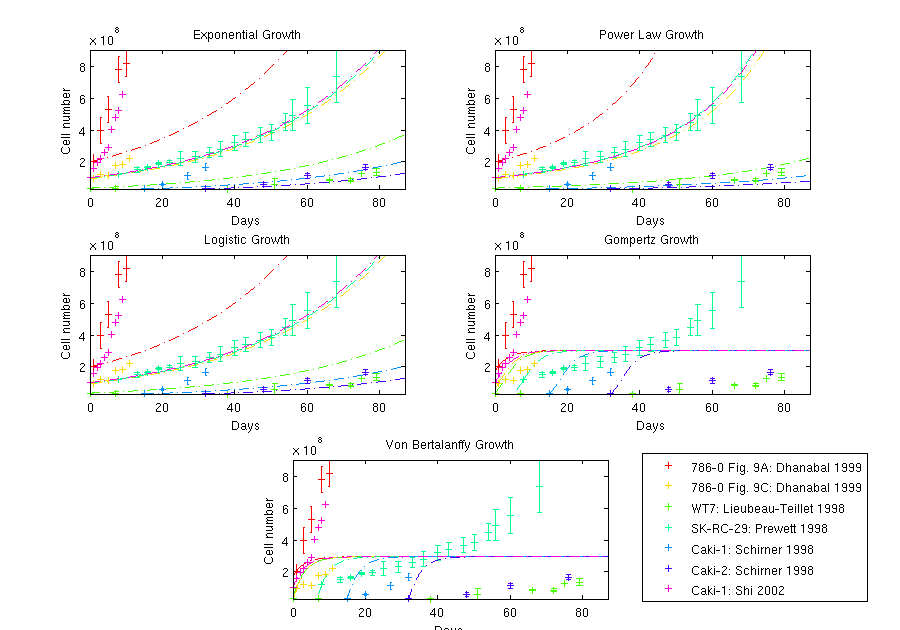}
\end{center}
\caption{Parameter Fitting to Combined \textit{In Vivo} Renal Cell Carcinoma Trials}
\end{figure}
\end{landscape}





\end{document}